\documentclass[twocolumn]{aastex631}

\defcitealias{Bochanski2011}{B11}

\shorttitle{Brown Dwarf Retrieval at Medium Spectral Resolution}
\shortauthors{Hood et al.}

\begin{document}

\title{Brown Dwarf Retrievals on FIRE!: Atmospheric Constraints and Lessons Learned from High Signal-to-Noise Medium Resolution Spectroscopy of a T9 Dwarf}

\correspondingauthor{Callie E. Hood}
\email{cehood@ucsc.edu}

\author[0000-0003-1150-7889]{Callie E. Hood}
\affiliation{Department of Astronomy \& Astrophysics, University of California,  Santa Cruz, CA 95064, USA}

\author[0000-0002-9843-4354]{Jonathan J. Fortney}
\affiliation{Department of Astronomy \& Astrophysics, University of California,  Santa Cruz, CA 95064, USA}

\author[0000-0002-2338-476X]{Michael R. Line}
\affiliation{School of Earth and Space Exploration, Arizona State University, Tempe, AZ 85287, USA}

\author[0000-0001-6251-0573]{Jacqueline K. Faherty}
\affiliation{Department of Astrophysics, American Museum of Natural History, Central Park West at 79th Street, NY 10024, USA}

\begin{abstract}

Brown dwarf spectra offer vital testbeds for our understanding of the chemical and physical processes that sculpt substellar atmospheres. Recently, atmospheric retrieval approaches have been successfully applied to low-resolution (R$\sim$100) spectra of L, T, and Y dwarfs, yielding constraints on the chemical abundances and temperature structures of these atmospheres. Medium-resolution (R$\sim 10^{3}$) spectra of brown dwarfs offer additional insight, as molecular features are more easily disentangled and the thermal structure of the upper atmosphere is better probed. We present results from a GPU-based retrieval analysis of a high signal-to-noise, medium-resolution (R$\sim$6000) FIRE spectrum from 0.85-2.5 $\mu$m of the T9 dwarf UGPS J072227.51-054031.2. At 60$\times$ higher spectral resolution than previous brown dwarf retrievals, a number of novel challenges arise. We examine the effect of different opacity sources, in particular for CH$_{4}$. Furthermore, we find flaws in the data like errors from order stitching can bias our constraints. We compare these retrieval results to those for a R$\sim$100 spectrum of the same object, revealing how constraints on atmospheric abundances and temperatures improve by an order of magnitude or more with increased spectral resolution. In particular, we can constrain the abundance of H$_{2}$S which is undetectable at lower spectral resolution. While these medium-resolution retrievals offer the potential of precise, stellar-like constraints on atmospheric abundances ($\sim$0.02 dex), our retrieved radius is unphysically small ($R=0.50_{-0.01}^{+0.01}$R$_{Jup}$), indicating shortcomings with our modeling framework. This work is an initial investigation into brown dwarf retrievals at medium spectral resolution, offering guidance for future ground-based studies and JWST observations.

\end{abstract}

\section{Introduction} \label{sec:intro}
Brown dwarfs, objects more massive than gas giant planets but yet not massive enough to sustain hydrogen fusion like a star \citep[13 M$_{Jup}$ $\lesssim$ M $\lesssim$ 73 M$_{Jup}$, ][]{Burrows2001}, provide essential testbeds of our understanding of the physics and chemical processes that sculpt substellar atmospheres. Without a sustained central energy source from fusion, brown dwarfs instead cool over time, leading to the formation of molecules and condensates in their atmospheres which dramatically affect their emitted spectra across the M, L, T, and Y spectral types \citep[e.g.][]{Kirkpatrick2005, Cushing2011}. The chemical and physical processes shaping these spectra are expected to be similar to those of gas giant exoplanets due to their similar effective temperatures \citep{Faherty2016}. Thus, the often more easily-observable spectra of brown dwarfs can inform our predictions for and interpretations of spectra of directly imaged planets. 

Traditionally, brown dwarf spectra have been compared to theoretical “grid models” which use our current understanding of substellar atmospheres and evolution to produce model spectra for a small number of fundamental parameters, such as composition, effective temperature, and surface gravity (see \citet{Marley2015} for a review). The cost of a small number of parameters is the number of chemical and physical assumptions, for example radiative-convective and thermochemical equilibrium, that are required. These grid models are an important resource for connecting observed properties with physical parameters of brown dwarfs as well as predicting signatures to be tested with future observations. However, while advancements in molecular opacities and the increasing model complexity have led to improved fits to observed spectra \citep{Phillips2020, Marley2021}, notable discrepancies remain \citep{Leggett2021} indicating there is still much to be learned about modeling these cool atmospheres. 

An alternative way to glean information from brown dwarf spectra is atmospheric retrieval, a data-driven Bayesian inverse method where minimal assumptions are made for the cost of far more free parameters. First developed for Earth and Solar System sciences \citep[e.g.][]{Rodgers2000, Fletcher2007} and then adapted for exoplanets \citep[e.g.][]{Madhu2009, Benneke2012, Line2013}, atmospheric retrievals have been applied successfully to brown dwarf spectra of various spectral types \citep{Line2014, Line2015, Line2017, Burningham2017, Burningham2021, Zalesky2019, Zalesky2022, Gonzales2020, Gonzales2021, Gonzales2022, Kitzmann2020, Piette2020, Howe2022, Lueber2022, Wang2022, Xuan2022, Calamari2022}. Retrievals provide a way to test the assumptions included in grid models; for example, \citet{Line2017} and \citet{Zalesky2019} used retrievals of T and Y dwarfs to show a  decrease of Na and K abundances with effective temperature, validating the rainout chemistry paradigm over pure equilibrium. However, while atmospheric retrieval can explore a wider range of possible atmospheres, unphysical combinations of parameters can still provide good fits to the data and therefore spuriously be preferred in retrieval frameworks. In particular, a number of brown dwarf retrieval studies have yielded unphysically small radius constraints \citep[e.g.][]{Burningham2021, Lueber2022} or very high surface gravities \citep{Zalesky2019}. Therefore, comparison to theoretical expectations from grid models are still needed to ensure retrieval results are fully contextualized. 

A vast majority of brown dwarf retrieval studies have been conducted on low-resolution (R $\sim$ 100) spectra. At medium-resolution, R $\gtrsim$ 1000, molecular bandheads are resolved into unique groups of densely-packed lines, allowing for more robust detections of molecules. Furthermore, the cores of strong lines are formed at lower pressures than can be sensed at low spectral resolutions, providing better probes of the upper end of the atmosphere’s temperature-pressure profile. Comparisons of medium-resolution spectra of brown dwarfs to grid models have provided validations of certain line lists \citep{Canty2015} and constraints on the brown dwarf’s fundamental properties \citep[e.g.][]{Bochanski2011, Petrus2022, Hoch2022}. Spectroscopy at medium-to-high spectral resolutions of brown dwarfs have been analyzed in retrieval frameworks, but often over a narrow wavelength range and with relatively low signal-to-noise, though they can also be combined with low-resolution observations for better constraints \citep{Wang2022, Xuan2022}. 

The aim of this work is to test how the atmospheric retrieval framework works at medium spectral resolution (R $\sim$ 6000), in terms of both new insights and novel challenges. We use the same framework successfully applied at low-resolution \citep[e.g.][]{Line2017, Zalesky2022} for a spectrum with roughly 60$\times$ higher spectral resolution. This work is structured as follows. In Section \ref{sec:methods}, we describe our dataset, retrieval framework, and the modifications necessary at this spectral resolution. In Section \ref{sec:results}, we give an overview of the tests performed and changes made throughout this project, and the accompanying results and lessons learned. In Section \ref{sec:discandcon}, we put our results in context with constraints from low-resolution spectra of the same object, previous analysis of this dataset, and grid models. Finally, our conclusions are summarized in Section \ref{sec:conclusion}.

\section{Methods} \label{sec:methods}

\subsection{Spectra of UGPS 0722} \label{subsec:specdata}
We perform our analysis on the medium-resolution spectrum of UGPS 0722 presented by \citet[hereafter \citetalias{Bochanski2011}]{Bochanski2011} obtained with the Folded-port InfraRed Echellette \citet[FIRE]{Simcoe2013} at the Magellan Telescopes. This spectrum covers 0.85 to 2.5 $\mu$m over 21 orders with R $\sim$ 6000. We scaled the reduced and order-stitched spectrum of \citetalias{Bochanski2011} to the observed H-band photometry on the Mauna Kea Observatories (MKO) photometric system following \citet{Line2017}. We will explore potential issues with which regions of this spectrum to include in our analysis (for example due to telluric absorption or order stitching problems) in later sections.

\begin{figure*}[!htbp]
\centering
\includegraphics[width=.95\linewidth]{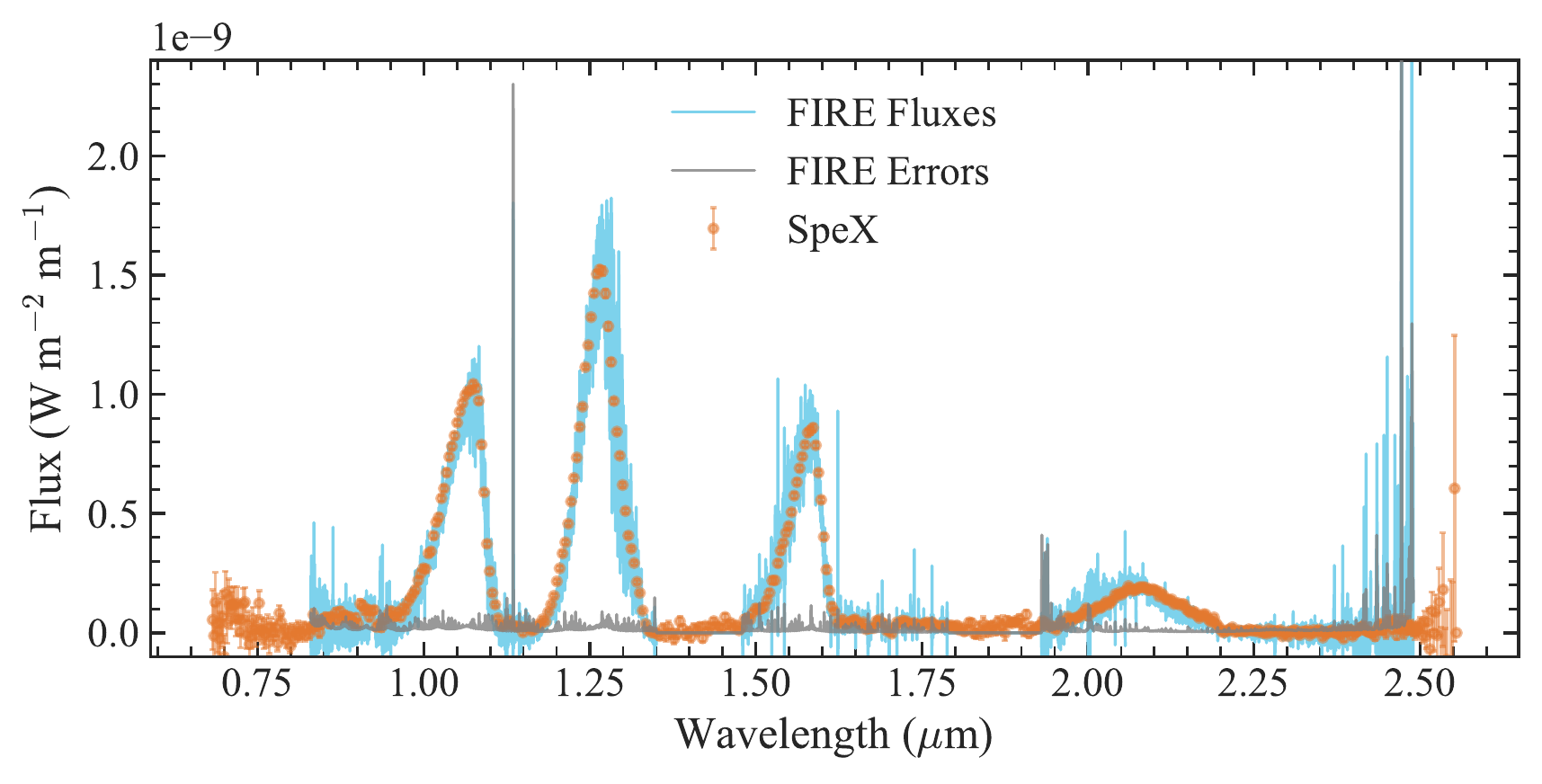}
  	\caption{The FIRE \citepalias[R $\sim$ 6000]{Bochanski2011} and SpeX \citep[R $\sim$ 87 - 300]{Burgasser2014} spectra of UGPS 0722. The FIRE fluxes are in blue, FIRE errors in gray, and SpeX datapoints in orange. The FIRE spectrum is available for download online as supplementary data.}
  	\label{fig:specdata}
\end{figure*}

We also use the low-resolution spectrum of UGPS 0722 from the SpeX Prism Library \citep{Burgasser2014}, which covers a similar wavelength range of 0.8 to 2.5 $\mu$m with a wavelength-dependent resolution of $\sim$ 87-300. This spectrum was also calibrated to flux units using the same H-band photometry as for the FIRE spectrum. The flux-calibrated FIRE and SpeX spectra are shown in Figure \ref{fig:specdata}. 

\subsection{GPU Retrieval Framework} \label{subsec:framework}

We use the CHIMERA retrieval framework successfully applied previously to low-resolution brown dwarf spectra \citep{Line2015, Line2017, Zalesky2019}. However, generating a forward model emission spectrum at a resolution of $\sim$ 60,000 (which is then binned to R $\sim$ 6,000) is quite computationally expensive. Thus, to make this study feasible, we must use a modified version for use with graphical processing units (GPUs), that builds upon the code described in \cite{Zalesky2022}. Specifically, we modify the radiative transfer to solve the two stream multiple scattering problem using the methods described in \cite{Toon1989}. However, as we are not in a particularly cloud regime, the effects of multiple scattering are neglible. As in the previous studies done with this framework, we use the affine-invariant MCMC ensemble sampler package \texttt{emcee} \citep{FM2013}.  We include uniform-with-altitude volume mixing ratios of H$_{2}$O, CH$_{4}$, CO, NH$_{3}$, H$_{2}$S, Na, and K, the surface gravity, a radius-to-distance scaling factor, the temperature-pressure (TP) profile, and three cloud parameters: cloud volume mixing ratio, the cloud pressure base, and the sedimentation efficiency \citep{Ackerman2001}. As in \cite{Line2017}, the TP profile is parameterized by 15 independent temperature-pressure points subject to two smoothing hyperparameters. These 15 temperature-pressure points are interpolated onto a finer 70 layer pressure grid for the radiative transfer using a cubic Hermite spline.

\begin{table}
\centering
\caption{Free Parameters in Our Retrieval Model} \label{tab:params}
\begin{tabular}{ll}
\tablewidth{0pt}
\hline
\hline
Parameter & Description \\
\hline
\decimals
log(f$_i$) & log of the uniform-with-altitude volume\\ 
& mixing ratios of H$_{2}$O, CH$_{4}$, CO, NH$_{3}$,\\
&   H$_{2}$S, Na, and K \\
log(g) & log surface gravity [cm s$^{-2}$] \\
(R/D)$^2$ & radius-to-distance scale [R$_{\rm Jup}$/pc]\\
T(P) & temperature at 15 pressure levels [K] \\
$b$ & errorbar inflation exponent \citep{Line2015}\\
$\gamma, \beta$ & TP profile smoothing hyperparameters\\
& \citep{Line2015}\\
log(Cloud VMR)   & log of the cloud volume mixing ratio \\ log(P$_{c}$)& log of the cloud base pressure \\ f$_{sed}$  & sedimentation efficiency \\
RV & radial velocity [km s$^{-1}$] \\
$v$ sin $i$ & rotational velocity [km s$^{-1}$] \\
\hline
\end{tabular}
\end{table}

\begin{table}
\centering
\caption{Opacity Sources for Our Retrieval Model} \label{tab:opacity}
\begin{tabular}{ll}
\tablewidth{0pt}
\hline
\hline
Species & Opacity Sources \\
\hline
\decimals
H2-H2, & \cite{Richard2012}\\ 
H2-He CIA & \\
H$_{2}$O & \cite{Polyansky2018} \\
CH$_{4}$ & (1) \cite{Yurchenko2014}\\
 & (2) \cite{Hargreaves2020} (Section \ref{subsec:medreslinelist})\\
CO & \cite{2010Rothman}, \\
 & isotopologues \cite{Li2015} \\
NH$_{3}$ & (1) \cite{Yurchenko2011}\\
 & (2) \cite{Coles2019} (Section \ref{subsec:medreslinelist}) \\
H$_{2}$S & \cite{Tennyson2012}, \\
& \cite{Azzam2015}, \\
 & isotopologues \cite{Rothman2013}\\
K & (1) see \cite{Marley2021}\\
 & (2) \cite{Allard2016} (Section \ref{subsec:alkali}) \\
Na & (1) see \cite{Marley2021} \\ 
 & (2) \cite{Allard2019} (Section \ref{subsec:alkali})\\ 
\hline
\end{tabular}
\end{table}

The chemical species and associated opacity sources used in this work are listed in Table \ref{tab:opacity}. We began with the set of absorption cross-sections presented in \cite{Freedman2008} and subsequently updated as detailed in \cite{Freedman2014}, \cite{Lupu2014}, and \cite{Marley2021}. However, we use a set of H$_{2}$O opacities calculated based on the POKAZATEL line list \citep{Polyansky2018}. In Section \ref{sec:results}, we explore the effect of switching our opacity source for a number of species, including NH$_{3}$ \citep{Coles2019}, CH$_{4}$ \citep{Hargreaves2020}, K \citep{Allard2016}, and Na \citep{Allard2019}. For the cloud opacity, we used Mie scattering theory assuming a Mg$_{2}$SiO$_{4}$ cloud with optical properties from \cite{Wakeford2015}. However, the exact cloud species assumed should not particularly matter as cloud optical properties tend to be gray over these near-infrared wavelengths and the cloud’s placement and extent in the atmosphere are parameterized independent of composition. 

At these moderate resolutions, the radial and rotational velocities present in the spectrum, thus we add these as two additional parameters in the retrieval forward model.  Both properties were measured in \citetalias{Bochanski2011} by comparison to grid models. We use the \textit{dopplerShift} function from \textsc{PyAstronomy} \citep{Czesla2019} to shift the forward modeled emission spectrum by a given radial velocity and interpolate it back onto the input wavelength grid. 

\begin{figure}
    \centering
    \includegraphics[width=.95\linewidth]{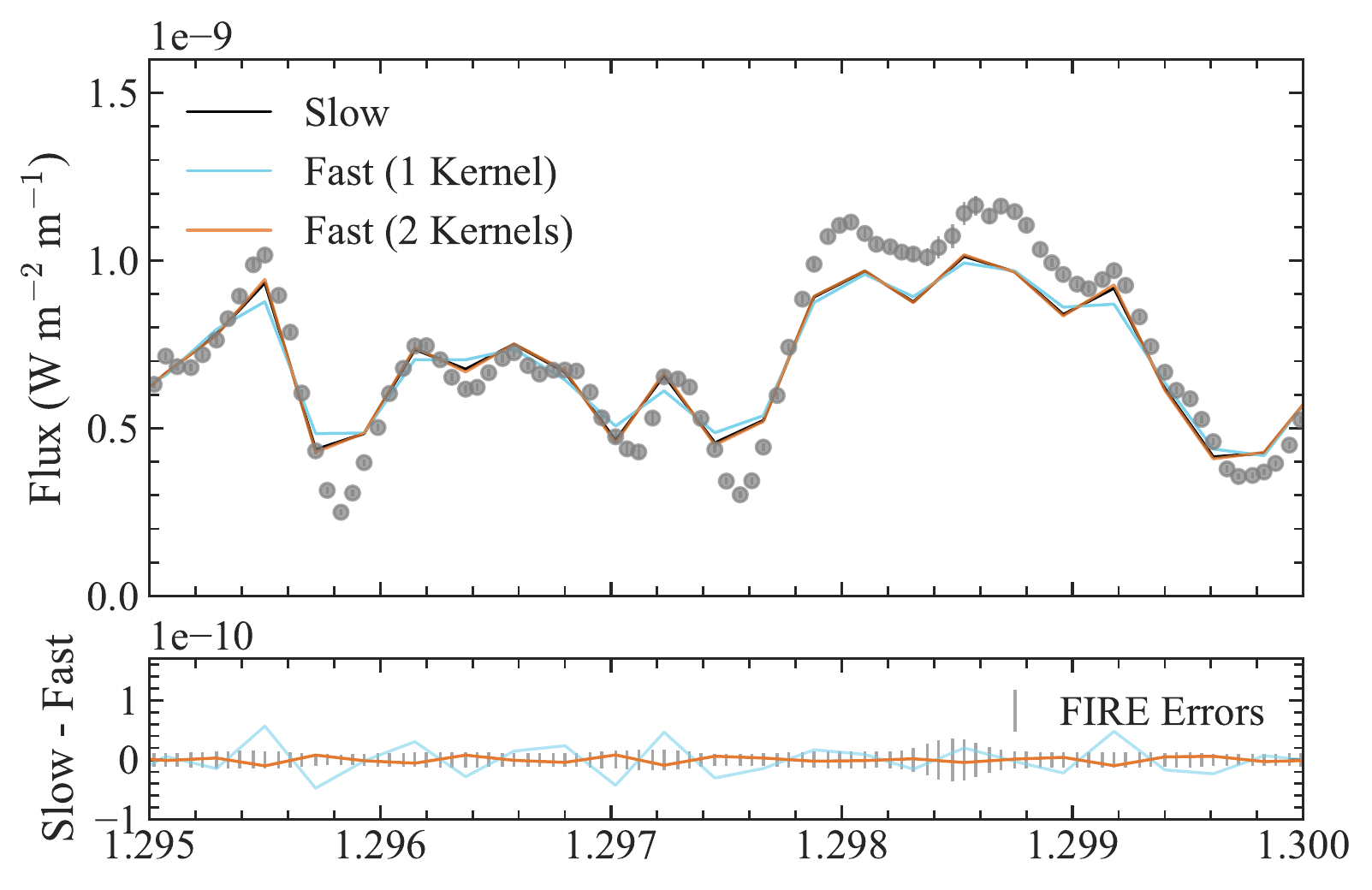}
  		\caption{Comparison of different rotational broadening methods. The top panel shows modeled emission spectra using different rotational broadening methods compared to a snippet of the FIRE spectrum of UGPS 0722 shown in the grey data points. The bottom panel shows the difference between the model broadened with a wavelength-dependent kernel (``Slow") and those using either 1 or 2 broadening kernels for the spectrum (``Fast"), and the data error bars in this region. At least two broadening kernels, using the median wavelengths of the blue and red halves of the spectrum, are required to reduce the difference with the slower, more accurate method to smaller than the error bars on the FIRE spectrum. }
  		\label{fig:broad}
\end{figure}

For the rotational velocity, we first tested the \textit{rotBroad} function from \textsc{PyAstronomy} which implements rotational broadening as described by \citet{Gray2008} for a given $v$ sin $i$ and linear limb-darkening coefficient (we chose 0). This function convolves the modeled spectral lines with a wavelength-dependent line profile representing the Doppler line broadening from rotation, called the ``broadening kernel." However, this method proved infeasibly slow, requiring 89 seconds to generate one broadened forward model spectrum. We then tested the \textit{fastRotBroad} function from \textsc{PyAstronomy}, which uses a single broadening kernel that only depends on the median wavelength of the input data, leading to a much faster forward model generating time of 0.19 seconds but differences from the slower version that were larger than the error bars of the FIRE spectrum as shown in Figure \ref{fig:broad}. As a compromise, we split our spectrum into two and used \textit{fastRotBroad} for each half, taking 0.2 seconds but leading to differences from the more accurate function that were smaller than the error bars. 

\begin{figure*}[!htbp]
    \centering
    \includegraphics[width=.95\linewidth]{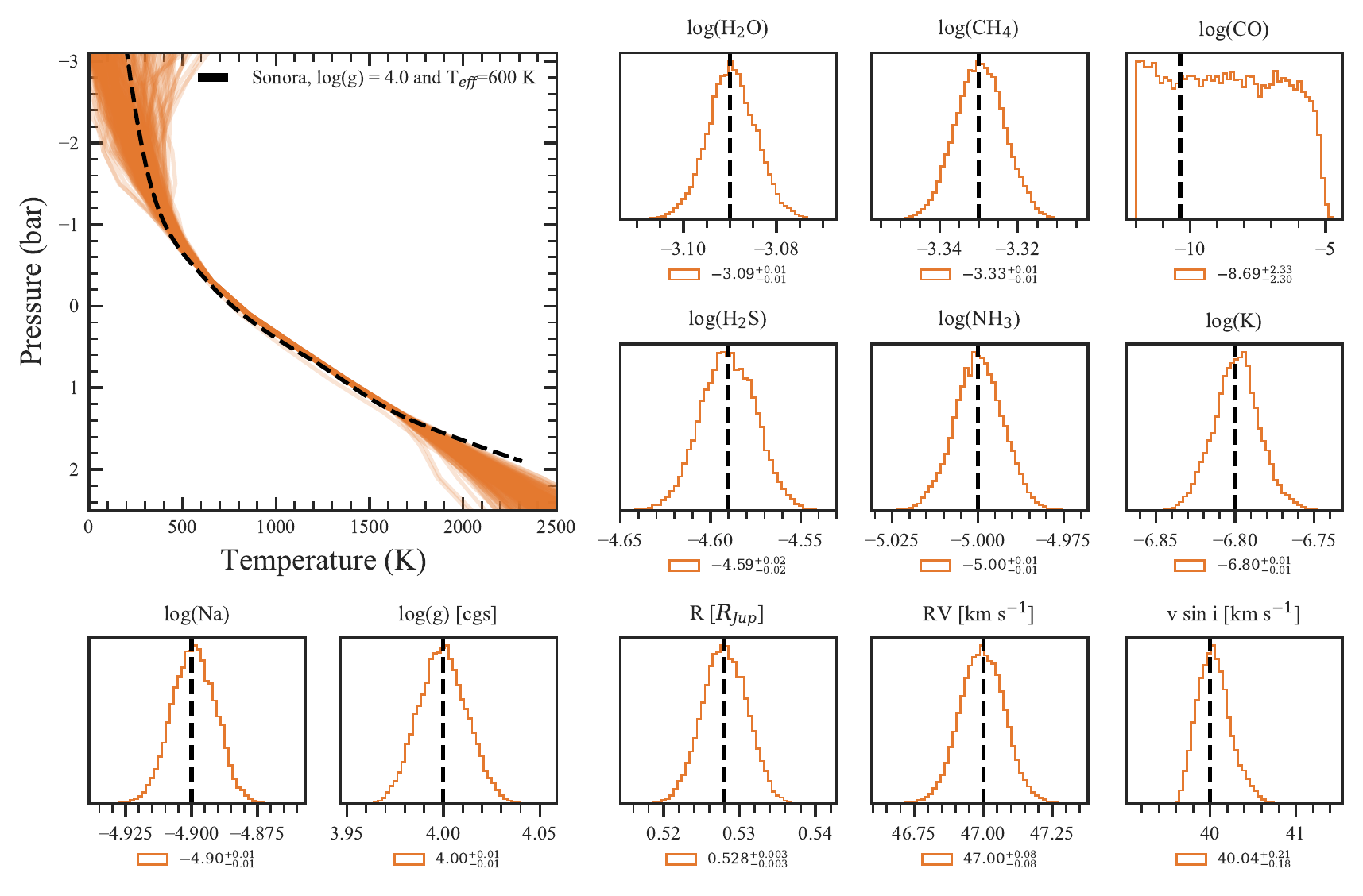}
  		\caption{Retrieved TP profiles and posteriors of certain parameters for a ``fake" FIRE spectrum based on the Sonora Bobcat model with an effective temperature of 600 K and surface gravity log(g)=4.0. }
  		\label{fig:SonoraFIRETest}
\end{figure*}

To test how our retrieval framework might perform on the FIRE data of UGPS 0722, we generated a fake test data set. We used CHIMERA to create one forward model based on the TP profile from the Sonora Bobcat grid \citep{Marley2021} for an object with T$_{eff}$ = 600 K and log(g)= 4.0, giving it a radial velocity of 47 km s$^{-1}$ and $v$ sin $i$ of 40 km $s^{-1}$.  We assume constant with altitude chemical abundances with log volume mixing ratios of H$_{2}$O = -3.09, CH$_{4}$ = -3.33, CO = -10.37, H$_{2}$S = -4.59, NH$_{3}$ =-5.00, K = -6.80, and Na = -4.90. This forward model was convolved to the FIRE instrument resolution, interpolated onto the wavelength grid of the FIRE spectrum, and given the same error bars as the FIRE spectrum of UGPS 0722. As such, we assume a $(R/D)^{2}$ value of 0.0158 (corresponding to a radius of $\sim$0.528 R$_{Jup}$ for an object at the distance of UGPS 0722) which is needed to give the same peak flux to error ratio as the FIRE spectrum of UGPS 0722. The results from a CHIMERA retrieval on this test dataset compared to the input values are shown in Figure \ref{fig:SonoraFIRETest}. We see that we are able to recover the input values with high accuracy and precision (with the exception of CO as it is unconstrained due to the low input mixing ratio), particularly with an unprecedented uncertainty of $\sim$ 0.01 dex on the chemical abundances and surface gravity, $\sim$10$\times$ more precise than constraints from spectra at R$\sim$100 \citep{Line2017, Zalesky2022}. 

\section{Results} \label{sec:results}
\subsection{Initial Fire Retrieval vs. SpeX}\label{sec:init}
Initially, we used the entire order-stitched and flux-calibrated FIRE spectrum UGPS 0722, with a few spurious data points effectively ignored by greatly inflating their error bars. However, in regions of high telluric absorption, the error bars on the spectrum were artificially underestimated (almost 14 orders of magnitude more precise than elsewhere). To prevent these data points from incorrectly driving our results, the error bars in these regions are inflated to a high enough value that these data points functionally do not contribute to our retrieval analysis. We also mask out those regions of strong telluric absorption in the SpeX spectrum as well to allow for a direct comparison. 

\begin{figure*}[htbp]
\centering
\includegraphics[width=.95\linewidth]{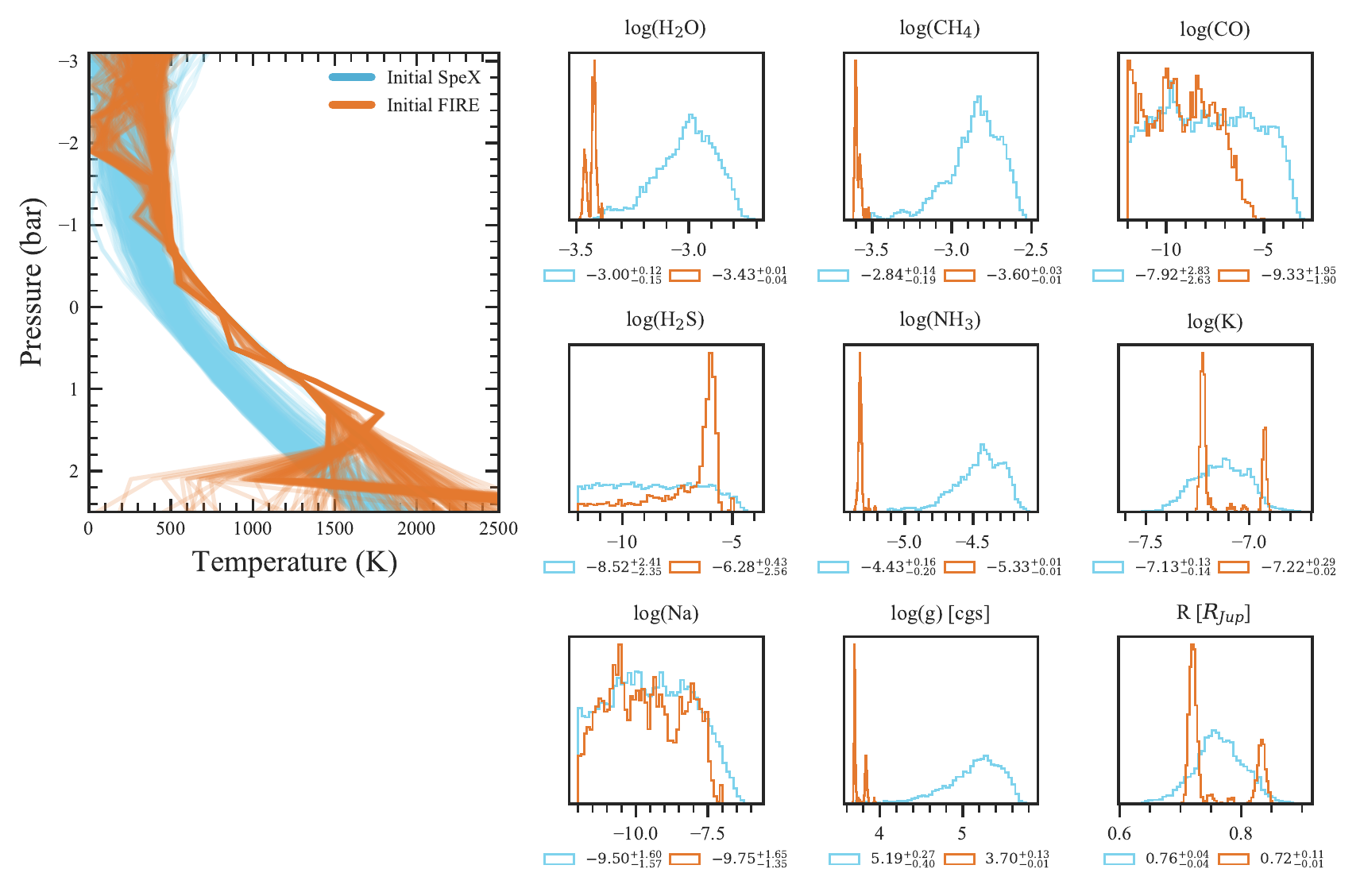}
\caption{Retrieved TP profiles and posterior distributions for selected parameters for initial retrievals on the SpeX and FIRE spectra of U0722.}
\label{fig:InitialSPEXvFIRE}
\end{figure*}

Figure \ref{fig:InitialSPEXvFIRE} shows the retrieved TP profiles and posterior distributions for selected parameters from our initial analysis on the UGPS FIRE spectrum compared to results from the SpeX spectrum.  Though the FIRE spectrum potentially offers more precision than the SpeX results, we find our MCMC chains have trouble converging, with bimodal posteriors, unphysical values for certain parameters like the surface gravity, and very jagged TP profiles. 

We suspected our forward model was too flexible, which was leading to overfitting and unphysical results. We reduced the number of parameters in our model by fixing the second TP profile smoothing hyperparameter $\beta$ and the cloud parameters to specific values. \cite{Line2017} found letting $\beta$ vary had a negligible effect compared to the nominal fixed value of $\beta = 5 \times 10^{-5}$ used by \cite{Line2015}. As previous work \citep{Line2015, Line2017} showed little evidence for optically thick clouds in T dwarfs, we set the cloud opacity parameters log(Cloud VMR), log(P$_{c}$), f$_{sed}$ = [-15, 2, 10], consistent with an optically thin cloud deck.

\begin{figure*}[htbp]
 \centering
\includegraphics[width=.95\linewidth]{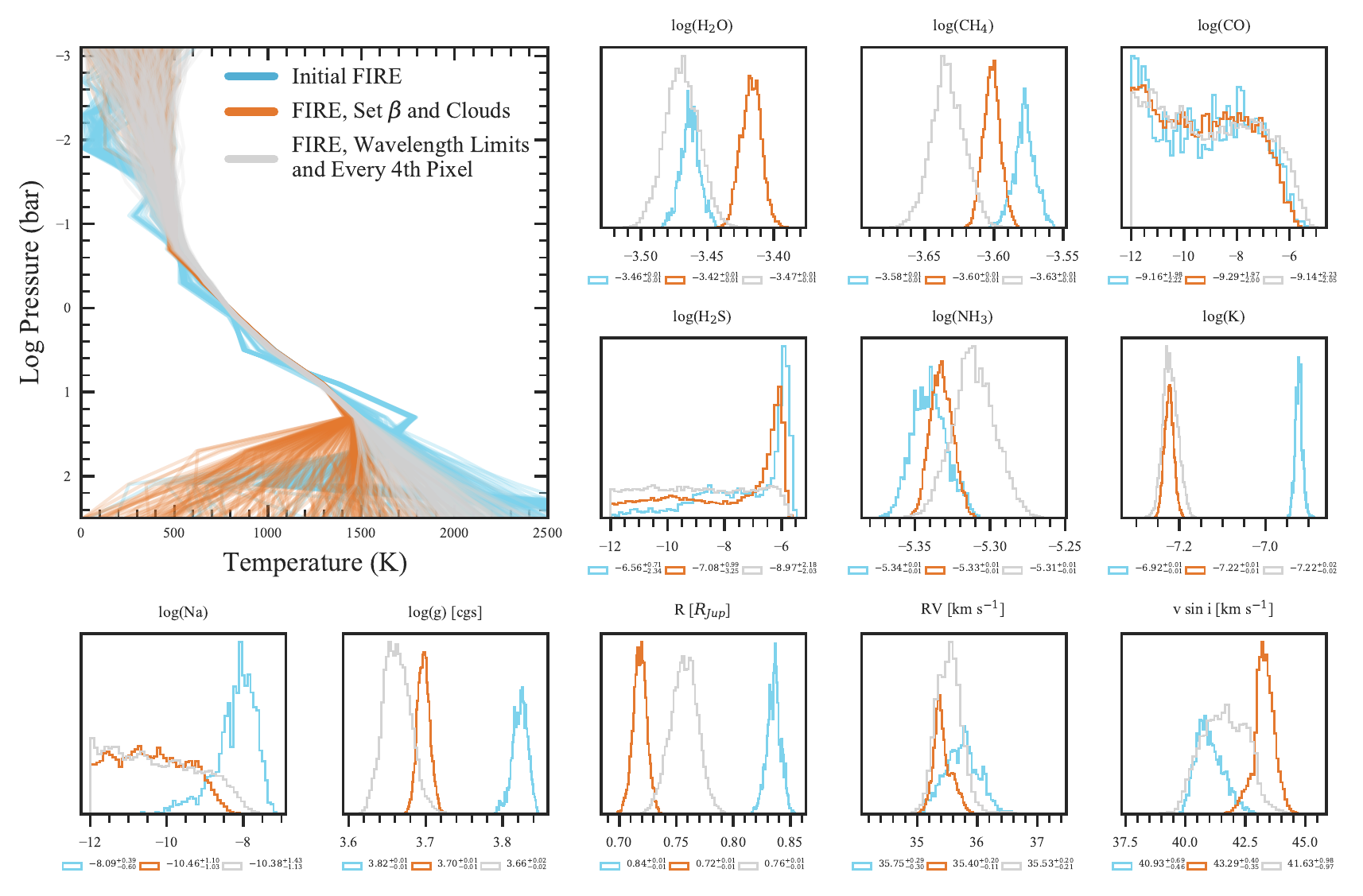}
\caption{Effect on the retrieved TP profiles and selected posteriors of fixing the smoothing hyperparameter $\beta$ and cloud parameters to set values (orange). In Section \ref{sec:resel}, we then limited the input spectrum to 0.9-2.35 $\mu$m and took every fourth pixel to limit the analysis to one data point per resolution element (grey).}
\label{fig:NoSmthorCld_every4}
\end{figure*}

The effect of fixing $\beta$ and the cloud parameters is shown in Figure \ref{fig:NoSmthorCld_every4} in orange. The TP profiles are smoother and more precisely constrained, while we no longer see bimodal posteriors for the plotted parameters. We note a potential detection of H$_{2}$S in this object, although the posterior on the H$_{2}$S abundance has a long lower tail. We will explore letting the cloud parameters vary again in Section \ref{sec:clouds}.

\subsection{Effect of Resolution Element and Wavelength Limits}\label{sec:resel}
Next, we decided to more mimic the analysis of \citetalias{Bochanski2011} and limit the FIRE spectrum to 0.9 - 2.35 $\mu$m when comparing to models, effectively getting rid of the noisiest regions of the spectrum at the beginning and end. Additionally, like SpeX, the FIRE spectrum is oversampled compared to a spectral resolution element, so we take every 4th pixel of the spectrum to ensure independent data points \citep[e.g.][]{Line2017, Kitzmann2020}.

Figure \ref{fig:NoSmthorCld_every4} shows the effect of these two changes in grey. While the retrieved TP profiles are very similar to the previous results at pressures less than $\sim$ 15 bars, the deep atmosphere is much warmer. The molecular abundances often shift slightly, but with the exception of H$_{2}$O the new posteriors are within 1$\sigma$. Our precision on almost all parameters also decreases, due to the substantial decrease in the number of included data points.

\subsection{Effect of Updated Line Lists}\label{sec:linelist}
In an effort to improve our best model fit to the observed spectrum, we investigated the effect of changing the sources of line lists used for CH$_{4}$ and NH$_{3}$. For NH$_{3}$, we upgraded to the more recent CoYuTe \citep{Coles2019} ExoMol line list instead of the older BYTe \citep{Yurchenko2011} list we were using previously. For CH$_{4}$, we replaced the ExoMol 10to10 list \citep{Yurchenko2014} with the recent HITEMP line list published by \cite{Hargreaves2020} which combined the more accurate ab initio line lists of \cite{Rey2017} with HITRAN2016 data \citep{Gordon2017}. 

\begin{figure*}[h]
    \centering
    \plottwo{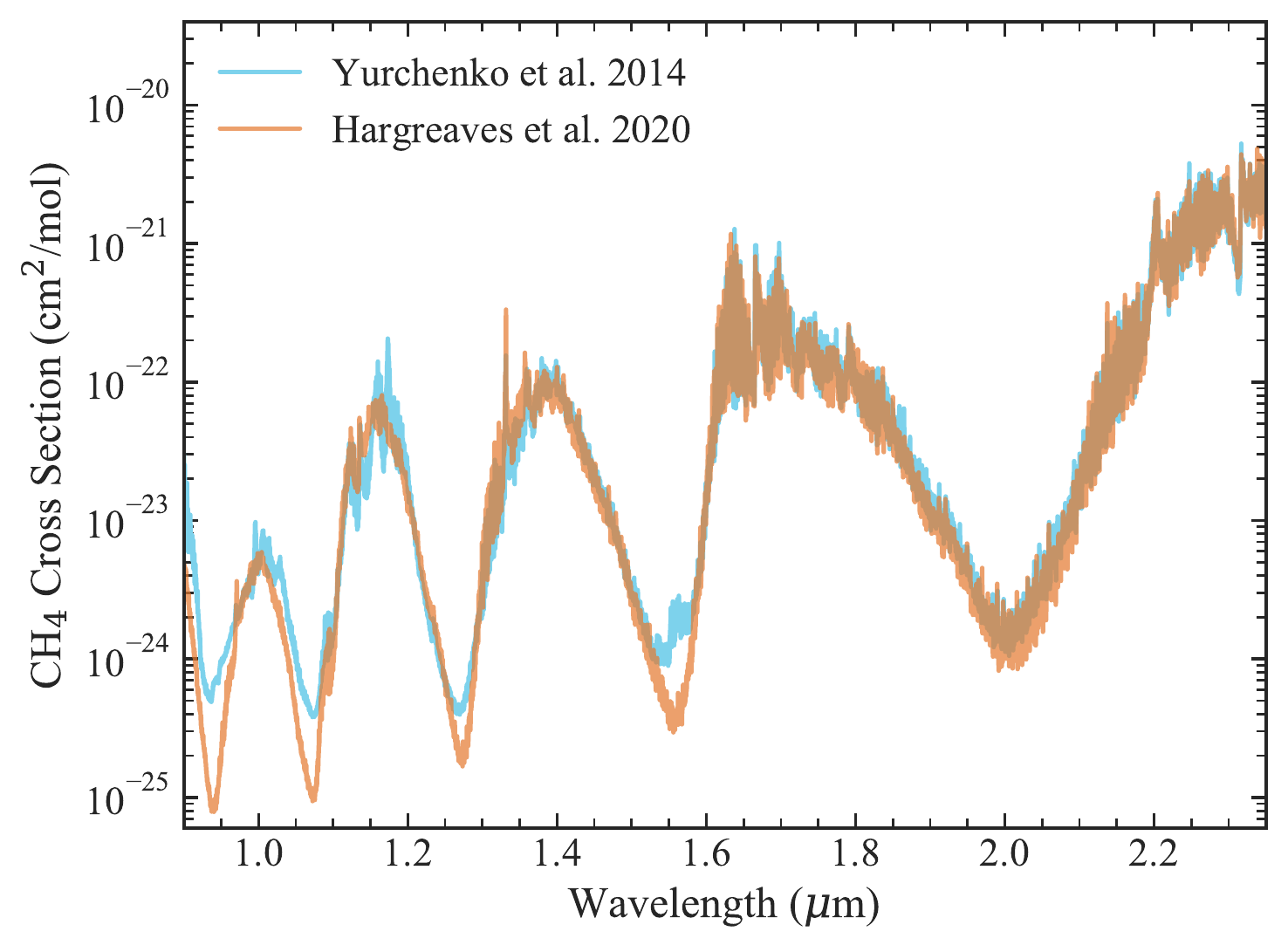}{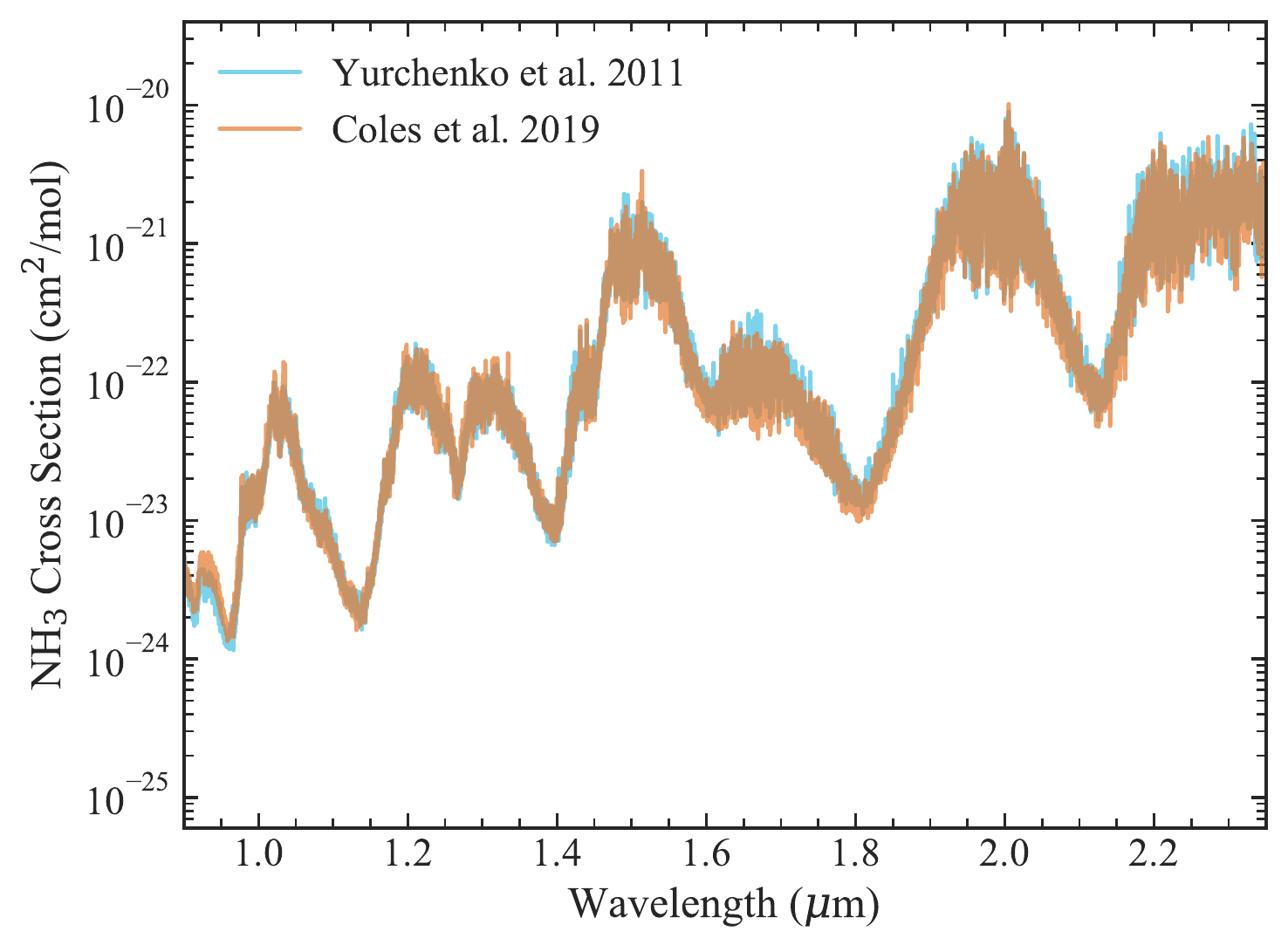}
  		\caption{Comparison of old and new molecular cross sections at 725 K and 1 bar, and smoothed to R $\sim$ 6000. \textit{Left}: Comparison of CH$_{4}$ cross sections from the ExoMol 10to10 \citep{Yurchenko2014} and HITEMP \citep{Hargreaves2020} line lists. \textit{Right}: Comparison of NH$_{3}$ cross sections from the older BYTe \citep{Yurchenko2011} and newer CoYuTe \citep{Coles2019} ExoMol line lists.}
  		\label{fig:CH4Xsec}
\end{figure*}

Comparisons of these ``old'' and ``new'' line lists for CH$_{4}$ and NH$_{3}$ at a specific pressure and temperature are shown in Figure \ref{fig:CH4Xsec}. The two CH$_{4}$ opacities on the left show clear deviations beyond just shifts to line positions, especially blueward of $\sim$ 1.62 $\mu$m. \cite{Hargreaves2020} demonstrate the better match of their CH$_{4}$ line list to experimental data for this wavelength region than the ExoMol CH$_{4}$ line list. In contrast, while minor differences are evident between the two NH$_{3}$ line lists in the right panel, they have similar overall values across the wavelength region of the spectrum. The effect of these changed line lists on our retrieved parameters is discussed below. 

\subsubsection{Low Resolution Retrievals}\label{sec:lowresret}
\begin{figure*}[htbp]
    \centering
    \includegraphics[width=.95\linewidth]{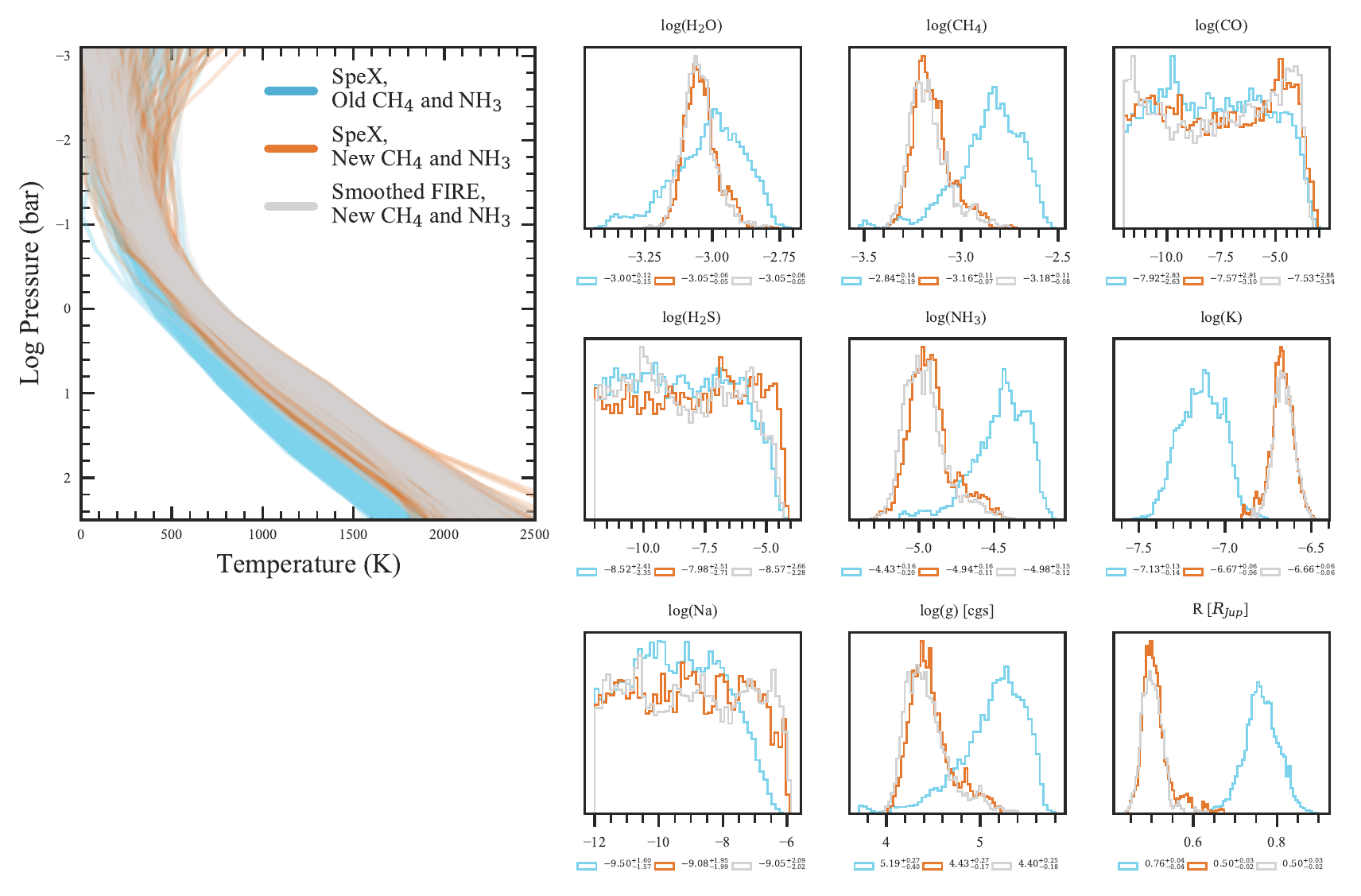}
  		\caption{Effect on the retrieved TP profiles and selected posteriors of updating the CH$_{4}$ and NH$_{3}$ line lists to those of \cite{Hargreaves2020} and \cite{Coles2019}, respectively, for the SpeX spectrum (orange). Results for the FIRE spectrum smoothed to the resolution of SpeX (grey) agree well for all parameters with those from the SpeX spectrum.}
  		\label{fig:SpeXNewXsec}
  	\end{figure*}

Figure \ref{fig:SpeXNewXsec} shows the effect of these updated line lists on the retrieved TP profile and selected posteriors for the SpeX spectrum. The retrieved TP profiles are consistently warmer than those obtained when using the \cite{Yurchenko2014} CH$_{4}$ and \cite{Yurchenko2011} NH$_{3}$ line lists. Constraints on the abundances of H$_{2}$O, CH$_{4}$, and NH$_{3}$ all shift to lower values, while the retrieved abundance of K increases by $\sim$ 0.5 dex. The retrieved surface gravity also decreases, from log(g)=5.19$_{{-0.40}}^{{+0.27}}$ to log(g)=4.43$_{{-0.17}}^{{+0.27}}$ (cgs). Furthermore, the retrieved radius, assuming the parallax distance of 4.12 $\pm$ 0.04 pc \citep{Leggett2012}, decreases from 0.76$_{{-0.04}}^{{+0.06}}$ to 0.5$_{{-0.02}}^{{+0.03}}$ R$_{Jup}$, an unphysically small value (see Section \ref{disc:spexvfire} for further discussion). Notably, the dramatic decrease in both surface gravity and radius are driven by updating the CH$_{4}$ line list, as this effect occurs even when the NH$_{3}$ line list is kept the same.

To assess if the FIRE dataset is consistent with the SpeX spectrum, we smoothed and sampled the FIRE data down to the SpeX resolution. Our retrieval results on this smoothed FIRE spectrum are compared to the original SpeX results (both with the updated cross sections) in Figure \ref{fig:SpeXNewXsec}. The retrieved constraints arising from the smoothed FIRE spectrum and the SpeX spectrum show remarkable agreement. This suggests that at low resolutions, observations at different times with completely different instruments produce extremely consistent results, given our retrieval model assumptions.

\subsubsection{Medium Resolution Retrievals}\label{subsec:medreslinelist}

\begin{figure*}[!htbp]
    \centering
    \plottwo{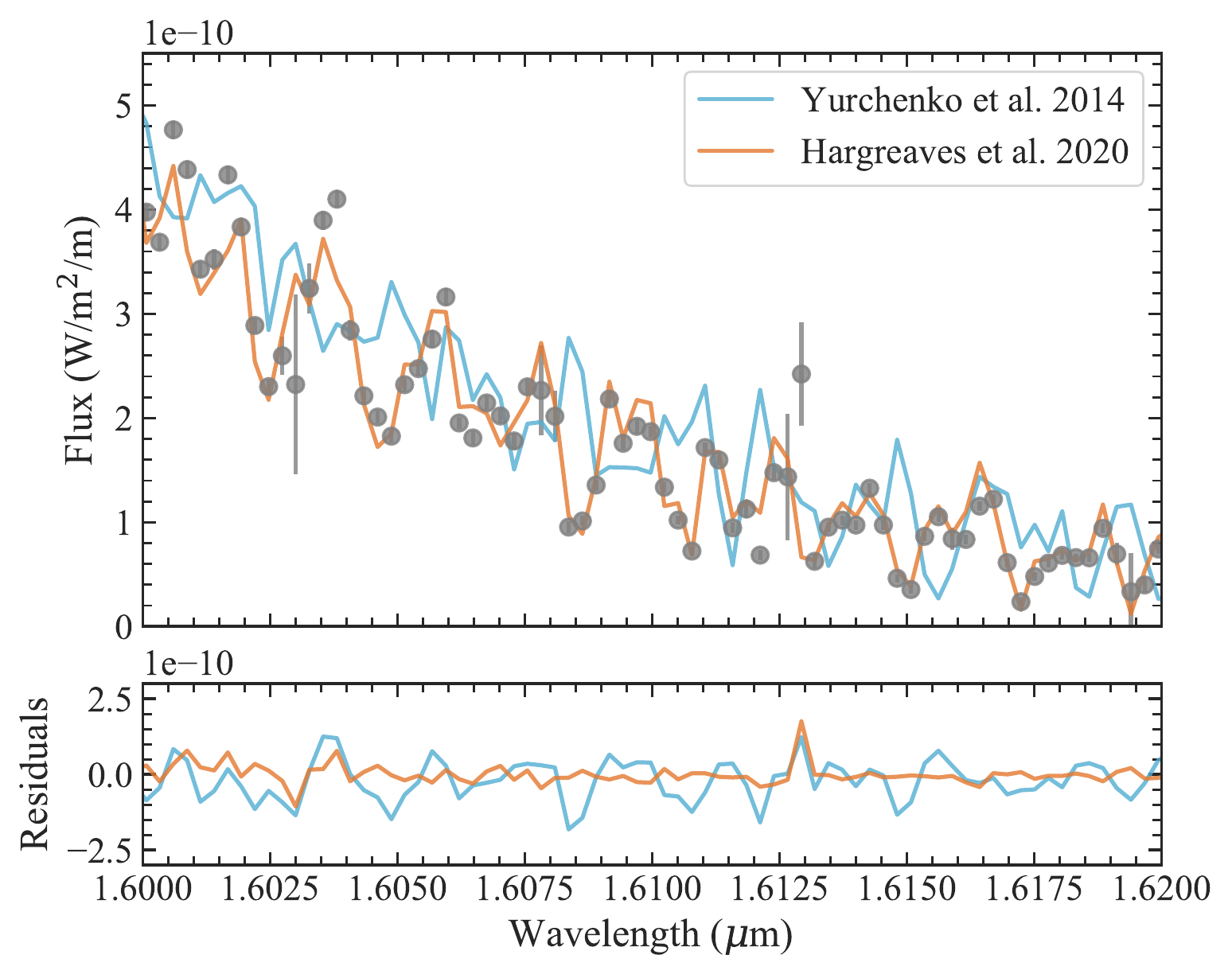}{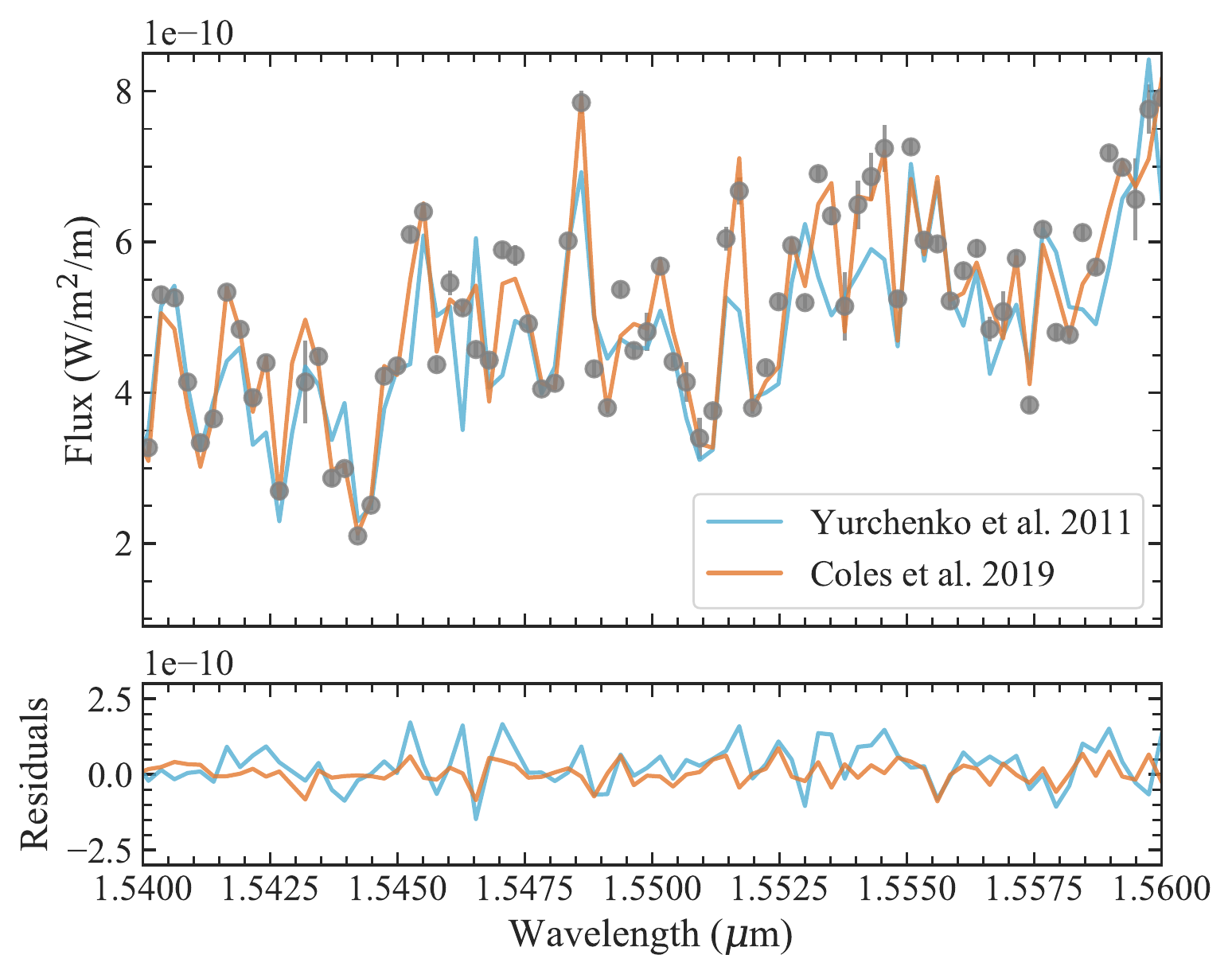}
  		\caption{Comparisons of median model spectra for retrieval results using old and new molecular line lists, compared to the FIRE spectrum of U0722 in narrow regions of the spectrum where CH$_{4}$ and NH$_{3}$ are expected to dominate on the left and right, respectively. \textit{Left:} Models with \cite{Yurchenko2014} and \cite{Hargreaves2020} CH$_{4}$ line lists; the \cite{Hargreaves2020} line list does a significantly better job at matching the CH$_{4}$ lines in this region. \textit{Right:} Models with \cite{Yurchenko2011} and \cite{Coles2019} NH$_{3}$ line lists; the \cite{Coles2019} line list improves the fit to NH$_{3}$ lines in this region.}
  		\label{fig:CH4ModelComp}
  	\end{figure*}

With the increased spectral resolution of FIRE, we can better assess the accuracy of the line positions in various line lists. Figure \ref{fig:CH4ModelComp} shows two narrow regions of the FIRE spectrum of U0722 where CH$_{4}$ and NH$_{3}$ are dominant in the left and right panels, respectively. For each molecule, we show model spectra generated with the updated line lists. In both cases, our newer line lists are better able to replicate the line positions of their respective molecules. Both the \cite{Hargreaves2020} and \cite{Coles2019} line lists incorporated empirical energy levels where available, as opposed to solely computed ones, leading to improved line position accuracy. 

\begin{figure*}[htbp]
         \centering
        \includegraphics[width=.95\linewidth]{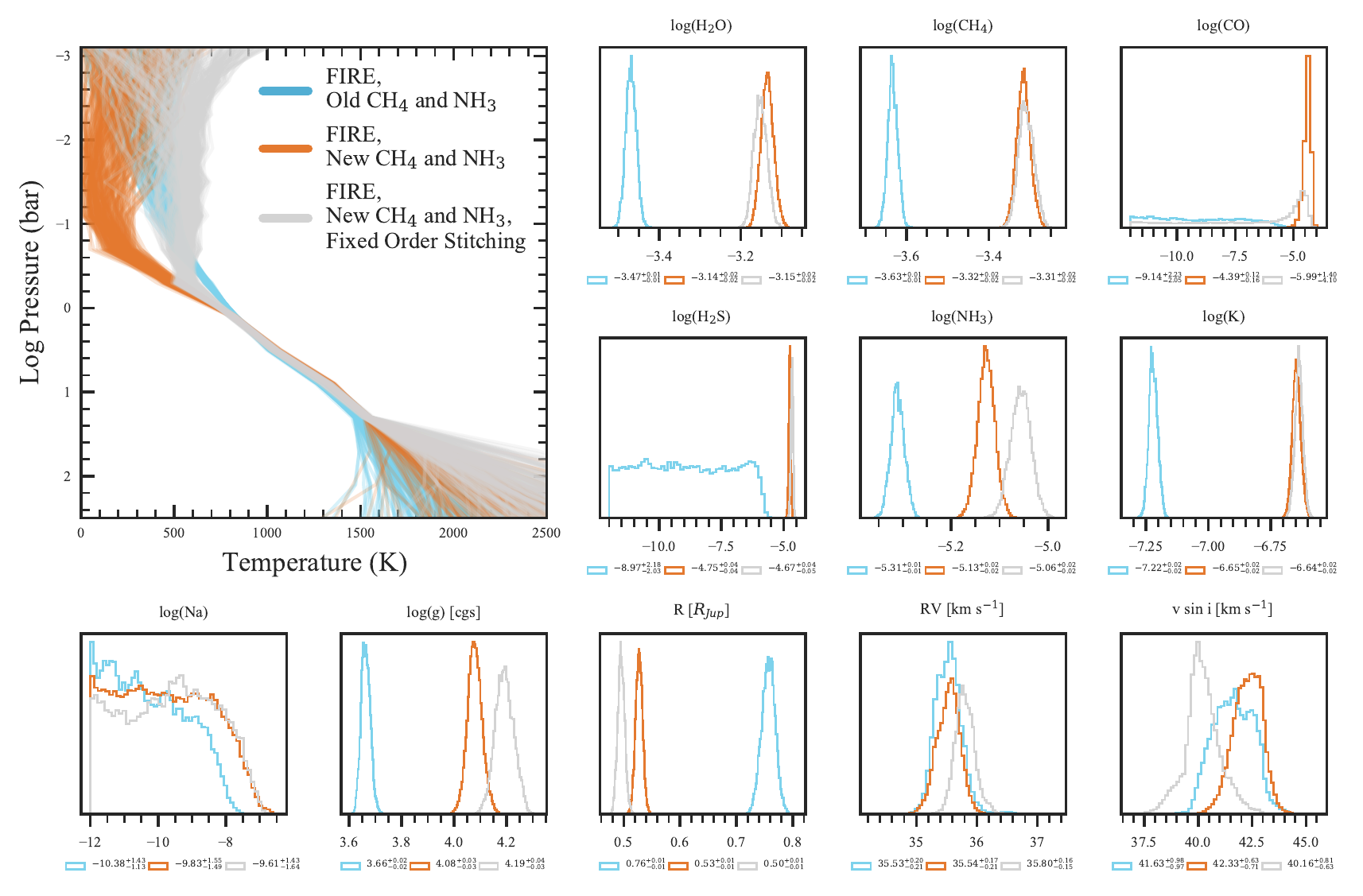}
  		\caption{Effect on the retrieved TP profiles and selected posteriors from the FIRE spectrum when updating the CH4 and NH3 line lists to those of \cite{Hargreaves2020} and \cite{Coles2019}, respectively (orange). The retrieved TP profiles and selected posteriors resulting from removing regions of the data where order stitching was not successful (Section \ref{sec:stitch}) are shown in grey.}
  		\label{fig:Corner_NewXsecs_stitch}
\end{figure*}

Unsurprisingly, improving the fit to line positions in our model does affect our retrieved atmospheric parameters. Figure \ref{fig:Corner_NewXsecs_stitch} shows how updating the line lists of CH$_{4}$ and NH$_{3}$ affects our retrieved TP profiles and posteriors in orange. While the TP profiles are relatively similar deeper than $\sim$ 1 bar, the scenario with updated line lists prefers the atmosphere to be as cold as possible around 0.2 bars. Our retrieved posteriors for all molecular abundances as well as potassium do shift to higher values. The surface gravity also increases, from a median log(g) = 3.66 $_{-0.02}^{+0.02}$  to log(g) = 4.08 $_{-0.03}^{+0.03}$(cgs), a more plausible value. However, the radius decreases just like for the SpeX retrieval in Section \ref{sec:lowresret}, from 0.76$_{-0.01}^{+0.01}$ to 0.53$_{-0.01}^{+0.01}$ R$_{Jup}$, an unphysically small size, which we discuss more in Section \ref{disc:spexvfire}. The retrieved radial velocity and $v$ sin $i$ are relatively unaffected by our choice of CH$_{4}$ and NH$_{3}$ line lists, indicating these measurements may be primarily driven by H$_{2}$O lines in the spectrum. 

\begin{figure}[!htbp]
         \centering
        \includegraphics[width=.95\linewidth]{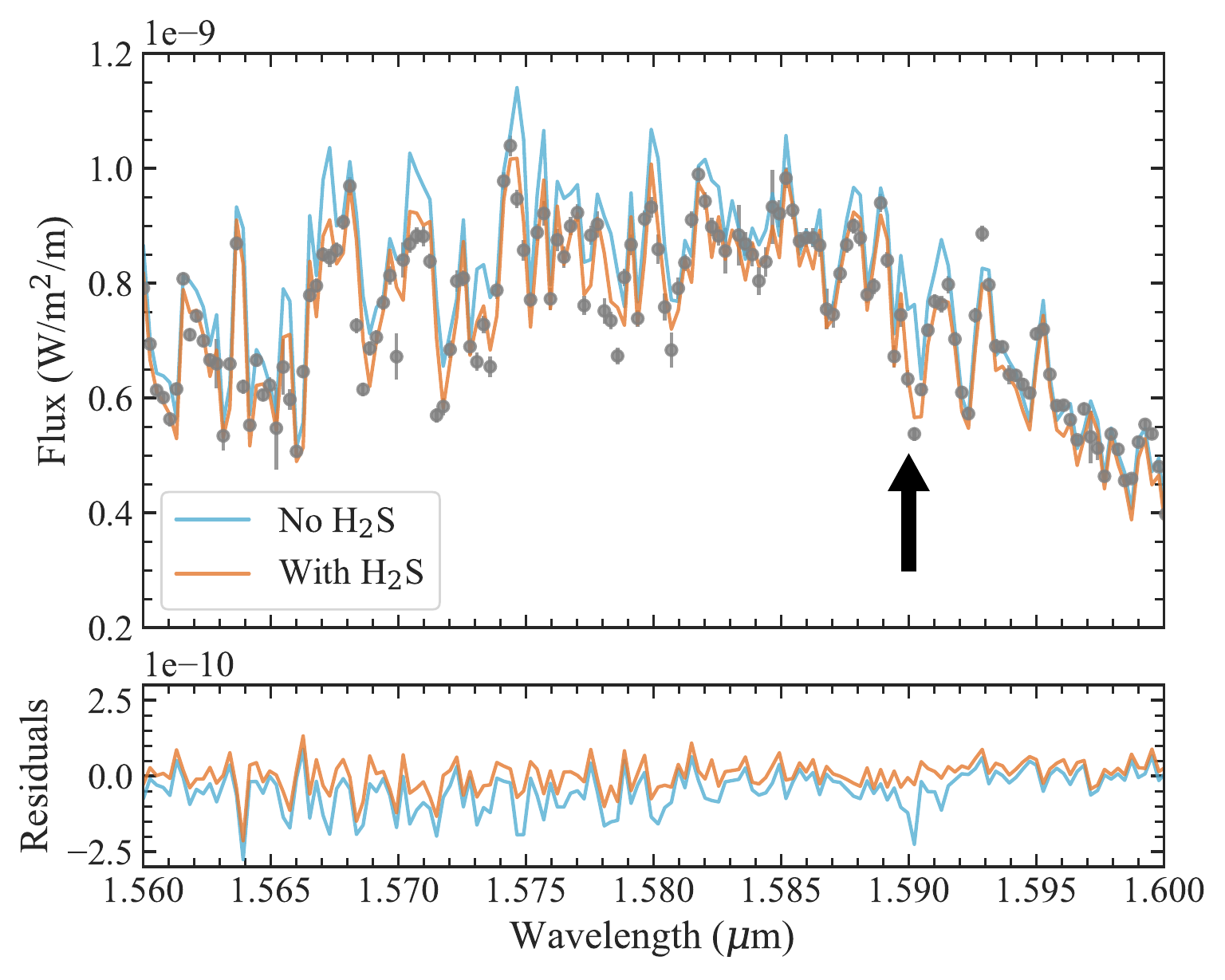}
  		\caption{Model spectrum from the median retrieved parameters generated with (orange) and without (blue) H$_{2}$S, compared to a snipppet of the FIRE spectrum where H$_{2}$S opacity is expected. Some features are better fit with the inclusion of H$_{2}$S. The black arrow indicates a blended H$_{2}$S line consistent with the H$_{2}$S detection of \cite{Tannock2022}.}
  		\label{fig:H2SModels}
\end{figure}

\begin{figure}[!htbp]
         \centering
        \includegraphics[width=.95\linewidth]{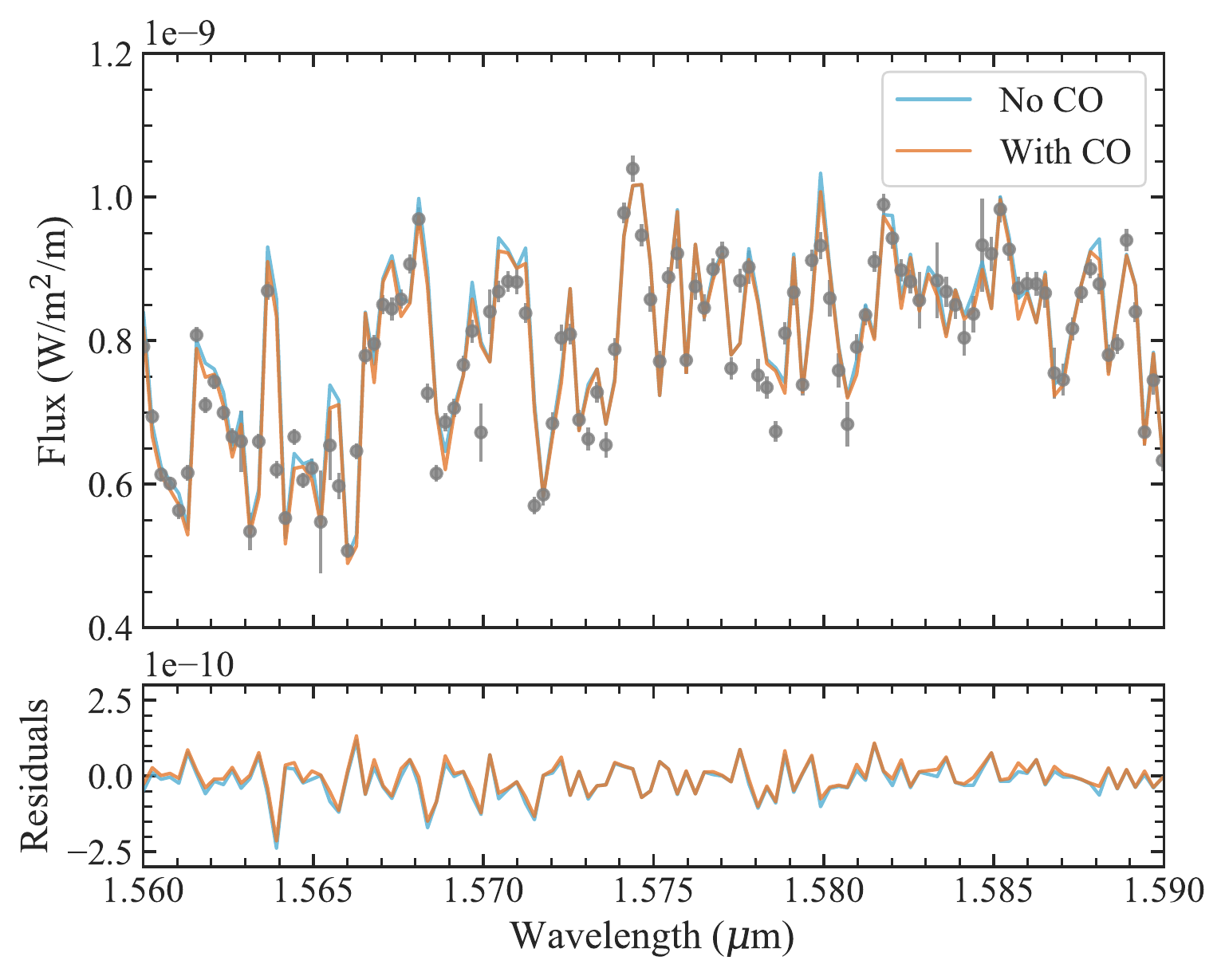}
  		\caption{Model spectrum from the median retrieved parameters generated with and without CO, compared to a snipppet of the FIRE spectrum where some CO opacity is expected. While including CO does slightly improve the fit to the data, the effect is slight.}
  		\label{fig:COModels}
\end{figure}

Notably, we retrieve bounded constraints on the H$_{2}$S and CO abundances; CO constraints were unbounded when using the older line lists. Figure \ref{fig:H2SModels} shows model spectra for the median retrieved parameters, with and without H$_{2}$S, where we expect H$_{2}$S opacity to have an effect. Both H$_{2}$S and CO have the greatest effect on our model spectra between $\sim 1.56 - 1.60 \mu$m where there is a window in the combined opacity of H$_{2}$O and CH$_{4}$. Including H$_{2}$S does improve the model fit in this region of the spectrum, but mostly as an overall shift in strength of features. However, around 1.59 $\mu$m there is what appears to be an H$_{2}$S line that is blended with another feature but clear in the residuals shown on the bottom. \cite{Tannock2022} reported an H$_{2}$S detection in  a high signal-to-noise, R $\sim$ 45,000 spectrum of a T6 dwarf, where they show a strong H$_{2}$S feature at this exact location around 1.59 $\mu$m. Thus, our constraint on the H$_{2}$S abundance is consistent with this feature, though it is dependent on a relatively small number data points ($\sim$3 points for the 1.59 $\mu$m feature and $\sim$ 20 others). Figure \ref{fig:COModels} shows models with and without CO, for the region of the spectrum where the model spectra are most different from each other. While the inclusion of CO does improve our model fit in this region, there is not a clear CO feature to point to as the source of our CO constraint. Thus, while we do have a bounded posterior for the CO abundance, it is perhaps a less trustworthy detection.


\subsection{Issues with Order Stitching}\label{sec:stitch}
\begin{figure}[h]
    \centering
    \includegraphics[width=.95\linewidth]{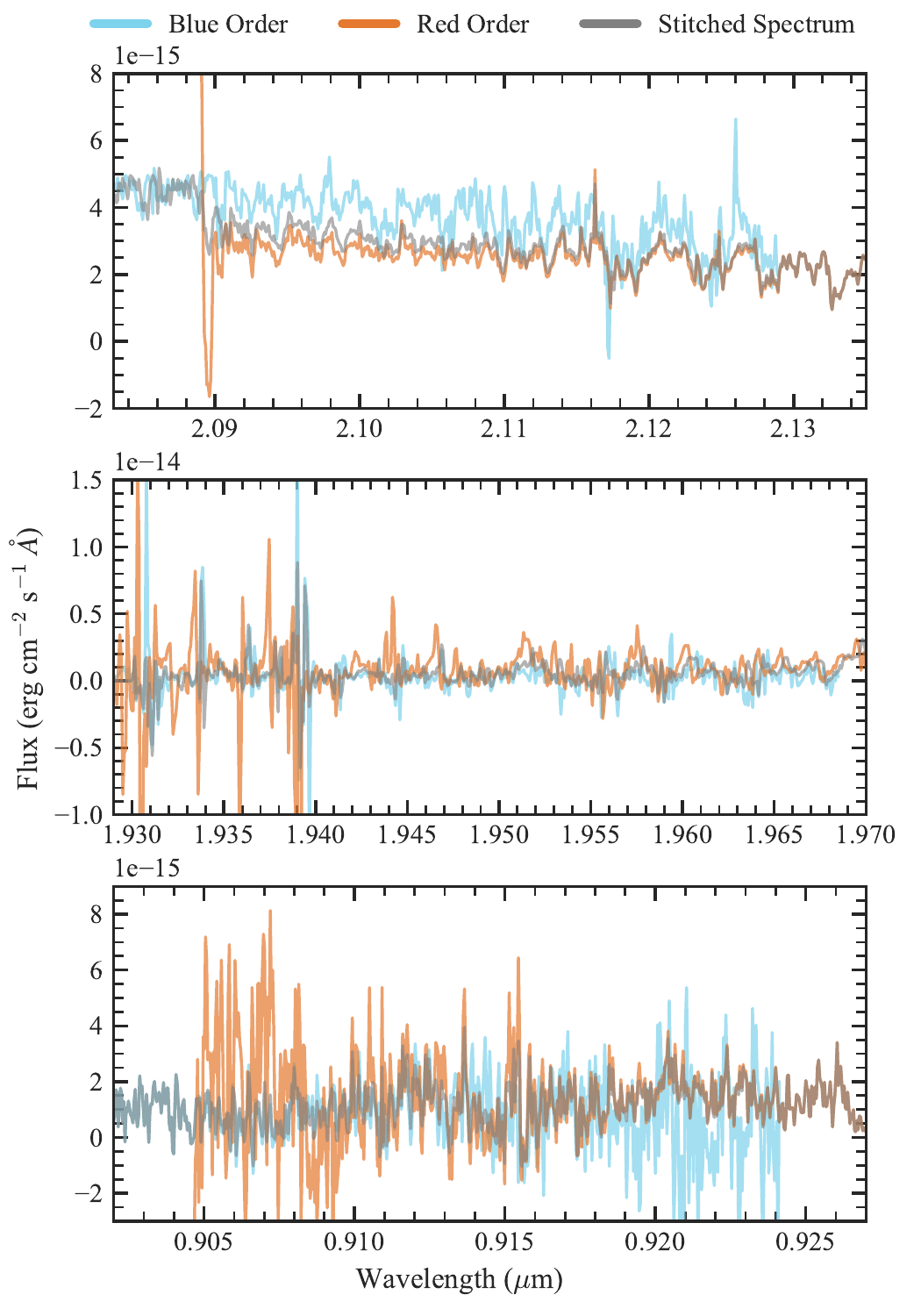}
  		\caption{Example order overlap regions of the stitched FIRE spectrum (grey) that were removed due to issues with order stitching, compared to the component orders (blue and orange).}
  		\label{fig:stitching}
\end{figure}

FIRE observations in cross-dispersed mode are spread over 21 different orders. These orders have some overlap in wavelength coverage on either end. The final stitched spectrum published by \cite{Bochanski2011} combined these orders into a single spectrum by averaging the regions where the orders overlapped. However, in some cases when one order was noisier than the other, this averaging can lead to data artifacts. We first noticed this issue when inspecting the spectrum around 2.1 $\mu$m, where a jump or step in the data appears that is not possible to reproduce with our forward models, as shown in the top panel of Figure \ref{fig:stitching}. Additional examples of overlap regions potentially subject to order stitching problems are shown in the lower two panels. While neither panel shows a step similar to the one seen at 2.09 $\mu$m, both cover regions where there is substantial disagreement between the two component orders leading to perhaps unphysical features in the stitched spectrum, particularly towards the middle of the overlapping areas. 

To avoid these potential order stitching issues from biasing our retrieved results, we artificially inflated the error bars in order overlap regions where the final stitched product differed from either input order spectrum by more than 10\%, effectively removing these data points from our analysis. \ref{fig:Corner_NewXsecs_stitch} shows the result of removing these order stitching artifacts on our retrieval results in grey. Though the region of the atmosphere shown in Figure \ref{fig:stitching} is near the peak of \textit{K} band and thus probes hotter parts of the atmosphere, other problematic order overlap regions masked by this process were on the edges of peaks in the spectrum and therefore contributed spuriously to constraints on the TP profile of the upper atmosphere. Our molecular and alkali abundances change slightly, with a less precise constraint on the amount of CO. While our median retrieved surface gravity increases to a slightly more plausible value of log(g)= 4.19 $_{-0.03}^{+0.04}$ from log(g) = 4.08 $_{-0.03}^{+0.03}$(cgs), our median radius shrinks even more to 0.50$_{-0.01}^{+0.01}$ R$_{Jup}$. Finally, our retrieved radial velocity and $v$ sin $i$ posteriors also slightly shift as well.

\subsection{Retrieving on Subsections of the Spectrum}\label{sec:partret}
While our stitched FIRE spectrum of U0722 covers \textit{y} - \textit{K} bands, we wanted to investigate the constraining power of different sets of spectral bands. In particular, \cite{Hargreaves2020} retains the completeness limits of the input line lists from \cite{Rey2017}, giving a temperature-dependent maximum wavenumber limit for which the CH$_{4}$ line list can be considered complete, or including all lines strong enough to affect the resulting opacity. This limit is 10700 cm$^{-1}$ or 0.93 $\mu$m for 1300 K, and 9500 cm$^{-1}$ or 1.05 $\mu$m for 1400 and 1500 K. Thus, the \cite{Hargreaves2020} CH$_{4}$ line list is not complete for the deeper, hotter temperatures probed by the \textit{y} band. Motivated by this potential completeness problem, we retrieved on the \textit{J} - \textit{K} band data only, followed by a retrieval on solely \textit{H} - \textit{K} bands. 

\begin{figure*}[htbp]
     \centering
    \includegraphics[width=.95\linewidth]{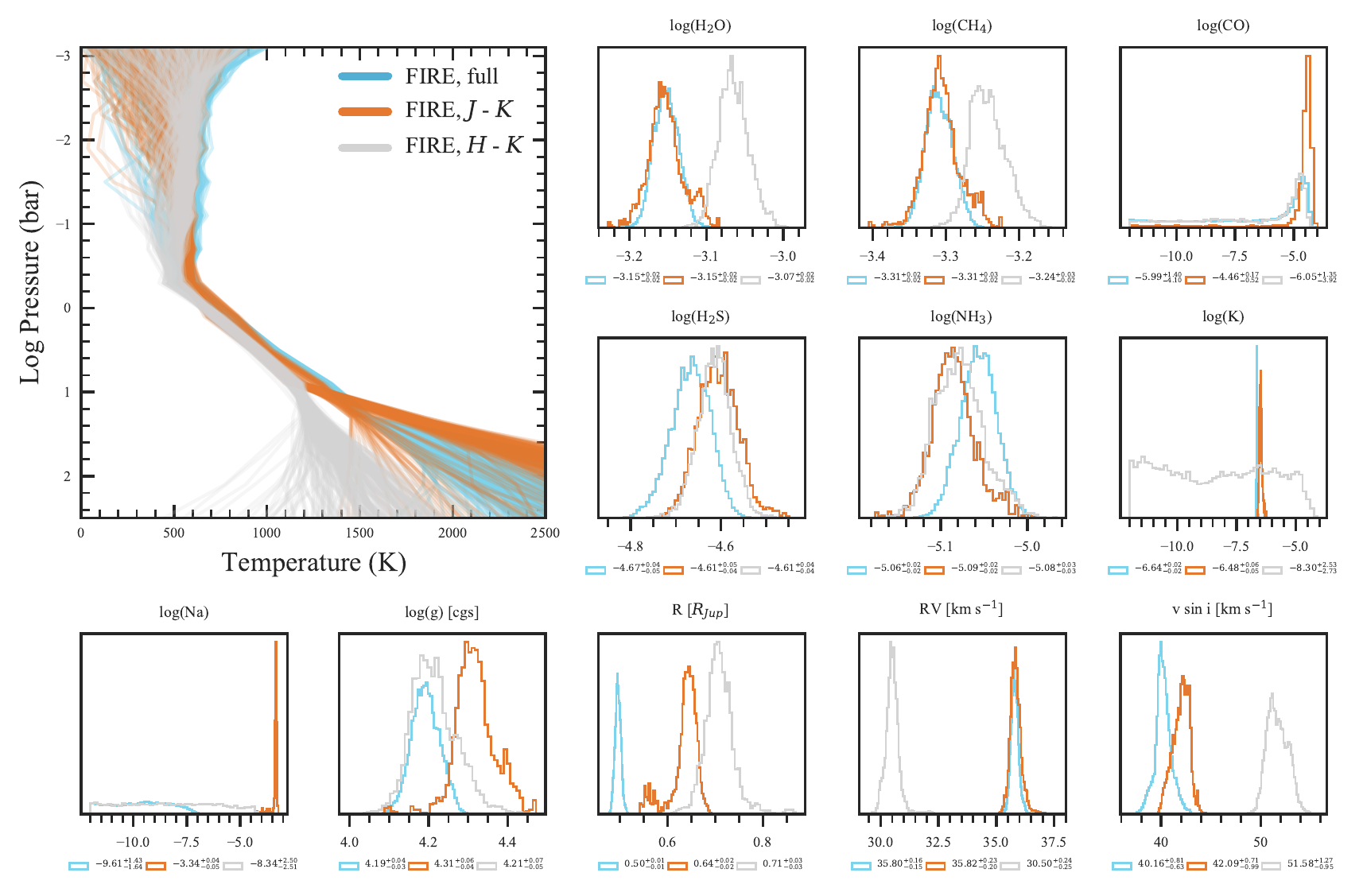}
  	\caption{Effect on the retrieved TP profiles and selected posteriors when performing the retrieval on the full wavelength range, \textit{J}-\textit{K}, or just \textit{H} and \textit{K} bands. }
  	\label{fig:DiffWavelengths}
\end{figure*}

Comparisons of the retrieved TP profiles and select posteriors are shown in Figure \ref{fig:DiffWavelengths}. The TP profiles get cooler as the shorter wavelength data is progressively removed from the analysis. As the flux in \textit{y} and \textit{J} bands comes from deeper in the atmosphere, we do not constrain the temperatures to as deep of pressures without these data, as indicated by the ``fanning out" of the TP profiles in these deeper layers. The H$_{2}$O and CH$_{4}$ abundances shift to higher values when just looking at \textit{H} and \textit{K} bands. While our constraint on the CO abundance becomes more precise for the \textit{J}-\textit{K} retrieval, there is still a long tail towards low values. The H$_{2}$S and NH$_{3}$ posteriors shift slightly with the exclusion of the \textit{y} band data. While we lose all constraints on the K abundance for just \textit{H} \& \textit{K}, we have a very precise and high constraint on the Na abundance when looking at \textit{J}-\textit{K} bands. Our retrieved surface gravity is also higher for the \textit{J}-\textit{K} retrieval than the other two, with a median of log(g)=4.31$^{+0.06}_{-0.04}$. We retrieve a larger radius as we remove each bluer band of data, perhaps reflecting the cooler TP profiles retrieved for these cases as well. Finally, removing the \textit{J} band data causes quite a shift in the retrieved radial velocity and $v$ sin $i$ values, pointing to the importance of the strong water lines in \textit{J} band to our constraints on these values for the full spectrum.

Given that the \cite{Hargreaves2020} line list is not complete for \textit{y} band at the temperatures found in this object, the \textit{J}-\textit{K} retrieval is perhaps a compromise between CH$_{4}$ line completeness while retaining some flux from deeper in the atmosphere. However, the anomalously high Na abundance does call into question the physical plausibility of these results.

\subsection{Effect of Alkali Opacities}\label{subsec:alkali}
\begin{figure}[htbp]
         \centering
        \includegraphics[width=.95\linewidth]{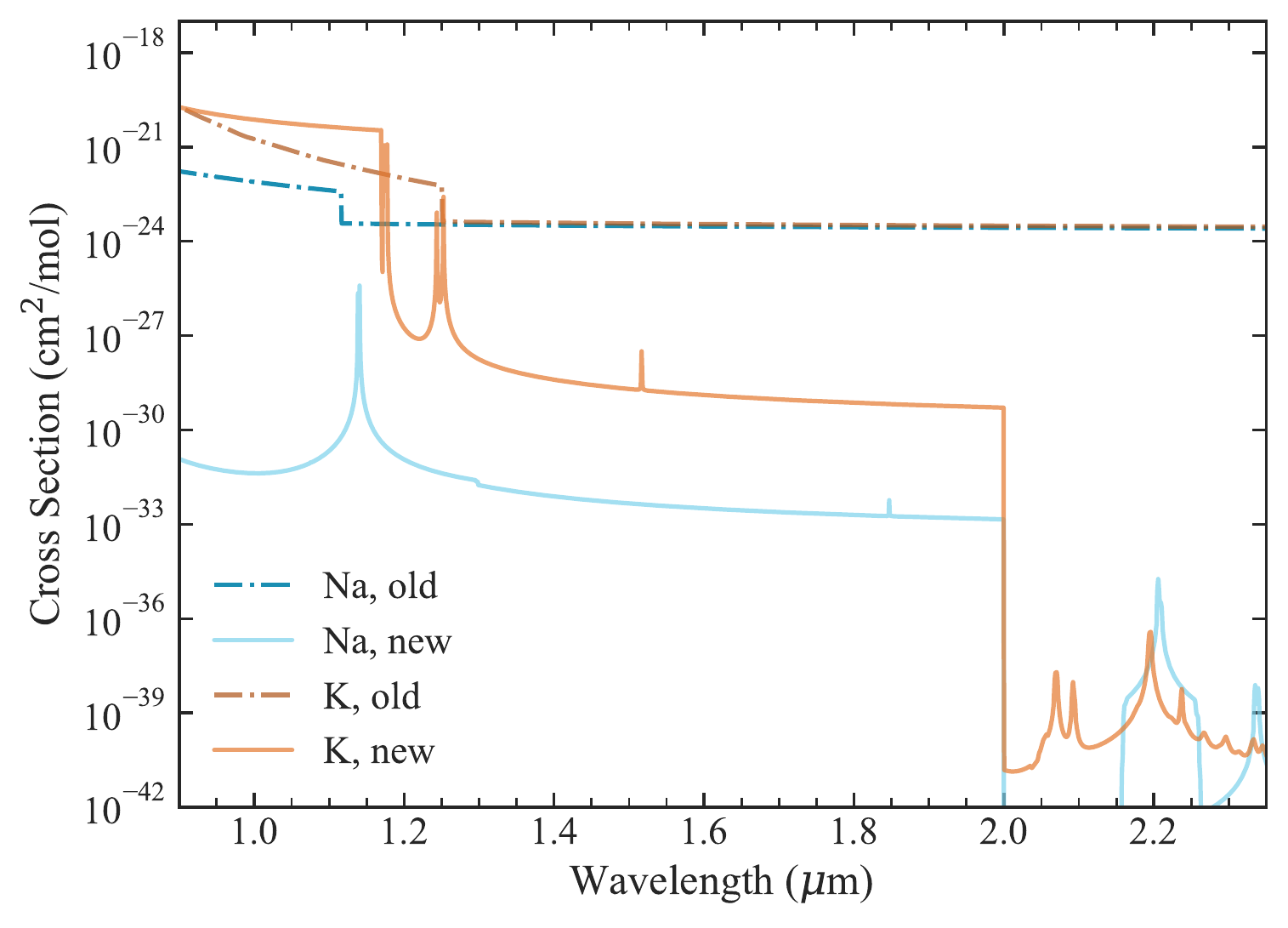}
  		\caption{Comparison of old (dash-dot) and new (solid) alkali cross sections at 725 K and 1 bar, and smoothed to R $\sim$ 6000. The older alkali cross sections are from \citet{Freedman2014}, while the newer ones are based on \citet{Allard2016} and \citet{Allard2019} for K and Na, respectively.}
  		\label{fig:Naxsec}
\end{figure}

Another change we made to our framework was to update the opacities used for the alkali metals Na and K. In particular, their strong lines at $\sim$~0.59 and 0.77 $\mu$m can be very broadened and significantly impact the near infrared spectrum of a brown dwarf. Prescriptions for these line profiles can vary quite a bit. We first used older alkali opacities based on the unified line-shape theory \citep{Allard2007a, Allard2007b}, as used in the Sonora Bobcat grid \citep{Marley2021}; example cross-sections of Na and K for 725 K and 1 bar are shown in Figure \ref{fig:Naxsec}, which basically become flat continuum opacity sources after a certain wing cutoff point. The high retrieved Na abundance for the \textit{J}-\textit{K} retrieval in Section \ref{sec:partret} above can then be understood as an additional source of continuum opacity used to reduce \textit{J} band flux which is not penalized when the \textit{y} band is excluded. These alkali opacities were shown by \cite{Gonzales2020} to produce more physically reasonable alkali abundances in their retrieval study of a d/sdL7+T7.5p binary than those from \cite{Burrows2003}. In contrast, newer cross sections based on the recent theoretical advancements of \cite{Allard2016, Allard2019} have more complicated shapes and significantly lower cross sections redward of $\sim$~1.2 $\mu$m. 

\begin{figure*}[htbp]
         \centering
        \includegraphics[width=.95\linewidth]{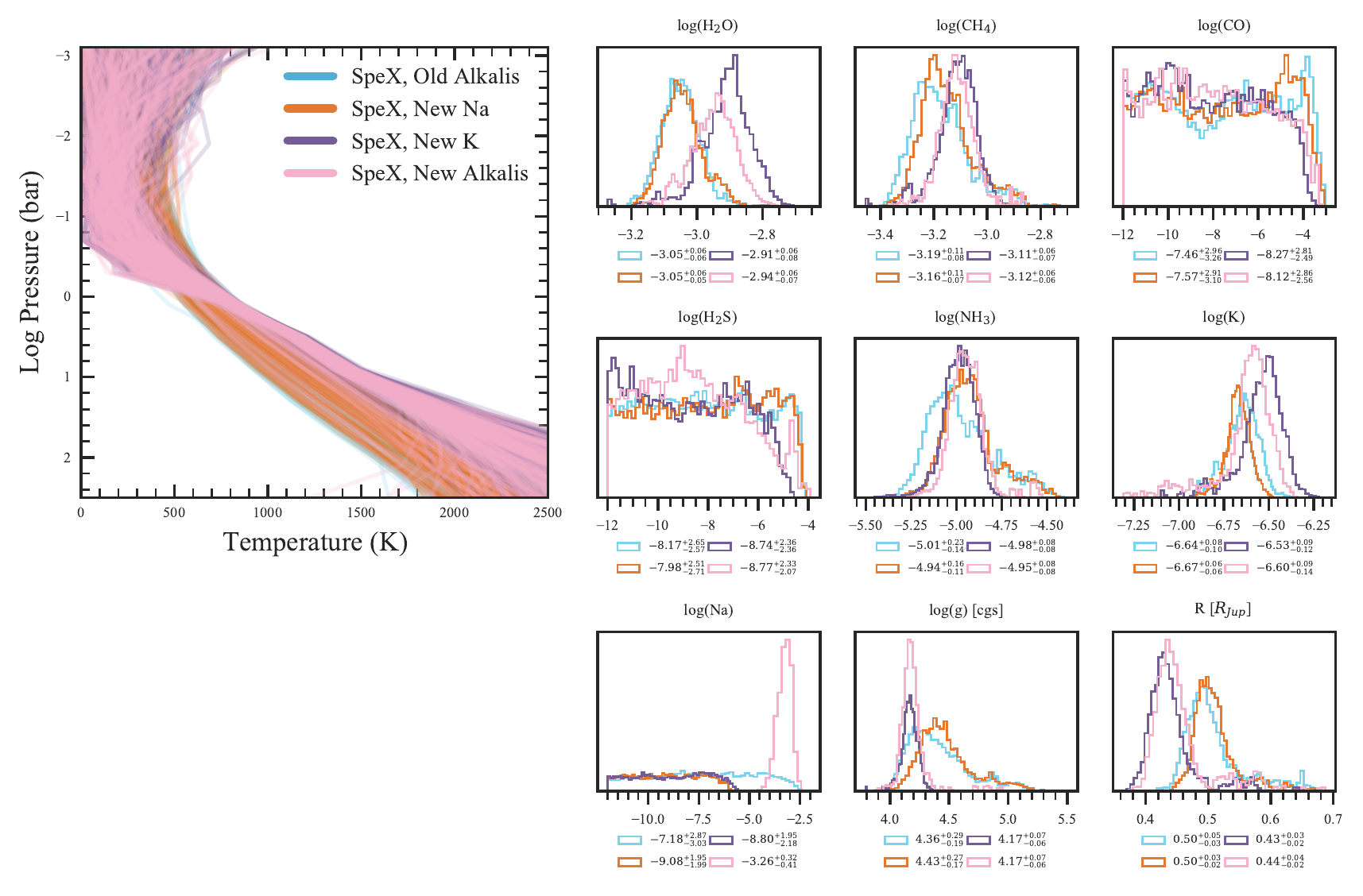}
  		\caption{Effect of updating alkali cross sections on retrieved TP profiles and selected posteriors using the SpeX spectrum.}
  		\label{fig:SpeX_newalk}
\end{figure*}

These differences demonstrated in Figure \ref{fig:Naxsec} between the older and newer line profiles can greatly affect our retrieved parameters. To isolate the differences, we retrieved on the SpeX data with just updating the Na cross sections, just updating K, and then updating both simultaneously. The retrieved TP profiles and selected posteriors for the different alkali treatments for the SpeX data are shown in Figure \ref{fig:SpeX_newalk}. Updating the Na opacity on its own does not particularly affect our retrieved results, perhaps unsurprisingly since the older K cross sections have so much more opacity. Updating to the new K causes a large change to our retrieved TP profiles, causing a very cold upper atmosphere with an inversion and a hotter, more precisely-constrained profile from $\sim$1 - 10 bars. The retrieved surface gravity and radius also decrease. Finally, updating Na and K at the same time is mostly similar to when just using the new K, except for a very high amount of retrieved Na.

\begin{figure*}[htbp]
         \centering
        \includegraphics[width=.95\linewidth]{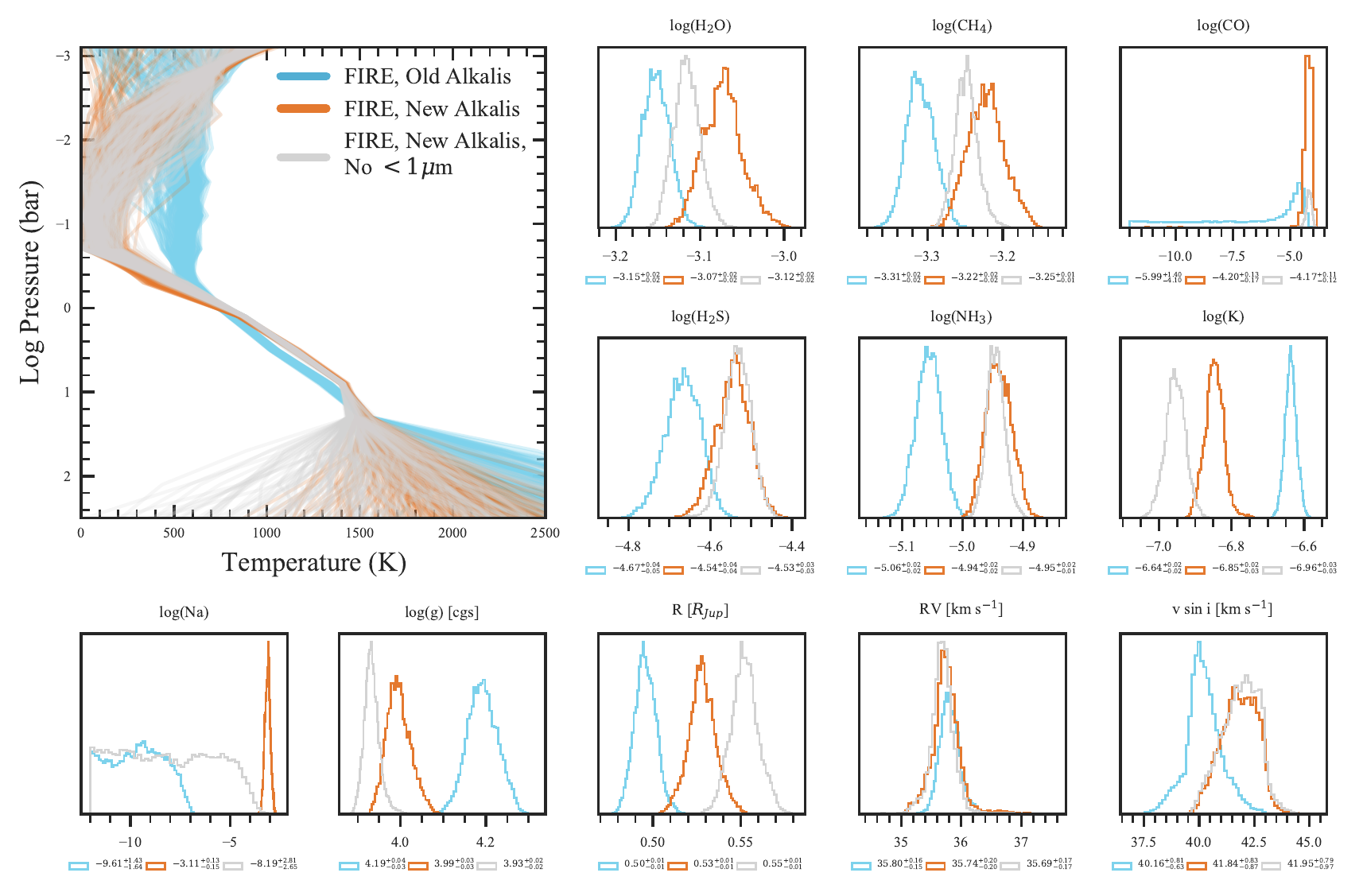}
  		\caption{Effect on the retrieved TP profiles and selected posteriors for the FIRE spectrum when updating our alkali cross sections (orange). As our Na constraint appears to be only from a small snippet on the blue end of the FIRE spectrum, removing any data $< 1 \mu$m leads to an unconstrained Na abundance (grey).}
  		\label{fig:FIRE_newalk_noblue}
\end{figure*}

Figure \ref{fig:FIRE_newalk_noblue} shows the effect of updating the alkali cross sections on our retrieved TP profiles and posteriors of selected parameters for the FIRE data of U0722. Similarly to the SpeX retrieval, the retrieved TP profile is hotter from $\sim$1 - 10 bars and goes to 0 K (unphysical) at $\sim$0.1 bars before inverting back to higher temperatures. The abundances of many species shift to higher values, particularly Na which again has an extremely high abundance. Additionally, the CO posterior loses its long low tail, leading to a higher median abundance.  We also retrieve a lower surface gravity, concordant with the SpeX results, while the radius increases slightly. 

\begin{figure}[htbp]
         \centering
        \includegraphics[width=.95\linewidth]{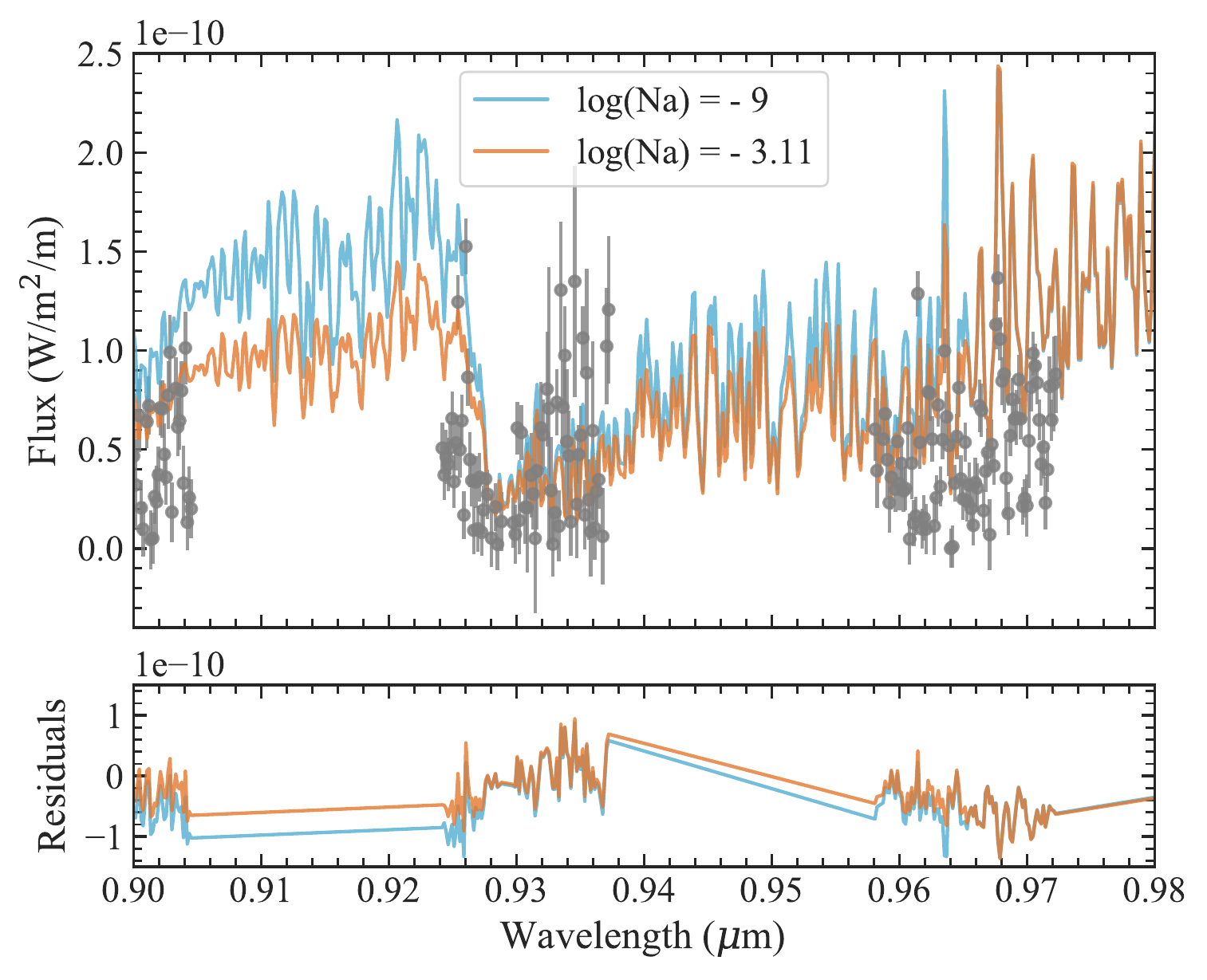}
  		\caption{Model spectra from the median retrieved parameters, both using the new alkali cross sections, generated with and without the retrieved high Na abundance. This wavelength region is the only one where the increased Na abundance leads to a notable difference in the model spectra.}
  		\label{fig:FIRE_highNa}
\end{figure}

Looking at a spectrum generated with the median retrieved parameters compared to one with significantly less Na, we see that the only discernible differences occur from $\sim$0.9 - 0.98 $\mu$m shown in Figure \ref{fig:FIRE_highNa}. The gaps in the FIRE data shown were regions where order stitching was found to be an issue as discussed in Section \ref{sec:stitch}. Although the high Na model does get closer to fitting the data in this region, it is still not a particularly good fit by eye, completely missing many of the data points. Furthermore, this is one of the noisiest regions of the data. Thus, the high retrieved Na abundance is likely a spurious result driven by low signal-to-noise data in these regions. 

We ran another retrieval on the FIRE data with the new alkali cross sections, but without any data blueward of 1 $\mu$m. The results are shown in Figure \ref{fig:FIRE_newalk_noblue}. The previous high Na constraint disappears as expected. While other posteriors shift as well, the constraints on the other species change only slightly. However, the inverted TP profile and even lower surface gravity compared to the retrieval with the old alkali cross sections are a cause for concern. Therefore, the old alkali cross sections are preferred as they give us more physical results. Further studies of the alkali line profiles and their effect on retrievals in particular would help provide context for these results and guidance for future medium resolution retrievals.

\subsection{Setting the Radius to 1 Jupiter Radius}\label{sec:radius}

\begin{figure*}[htbp]
         \centering
        \includegraphics[width=.95\linewidth]{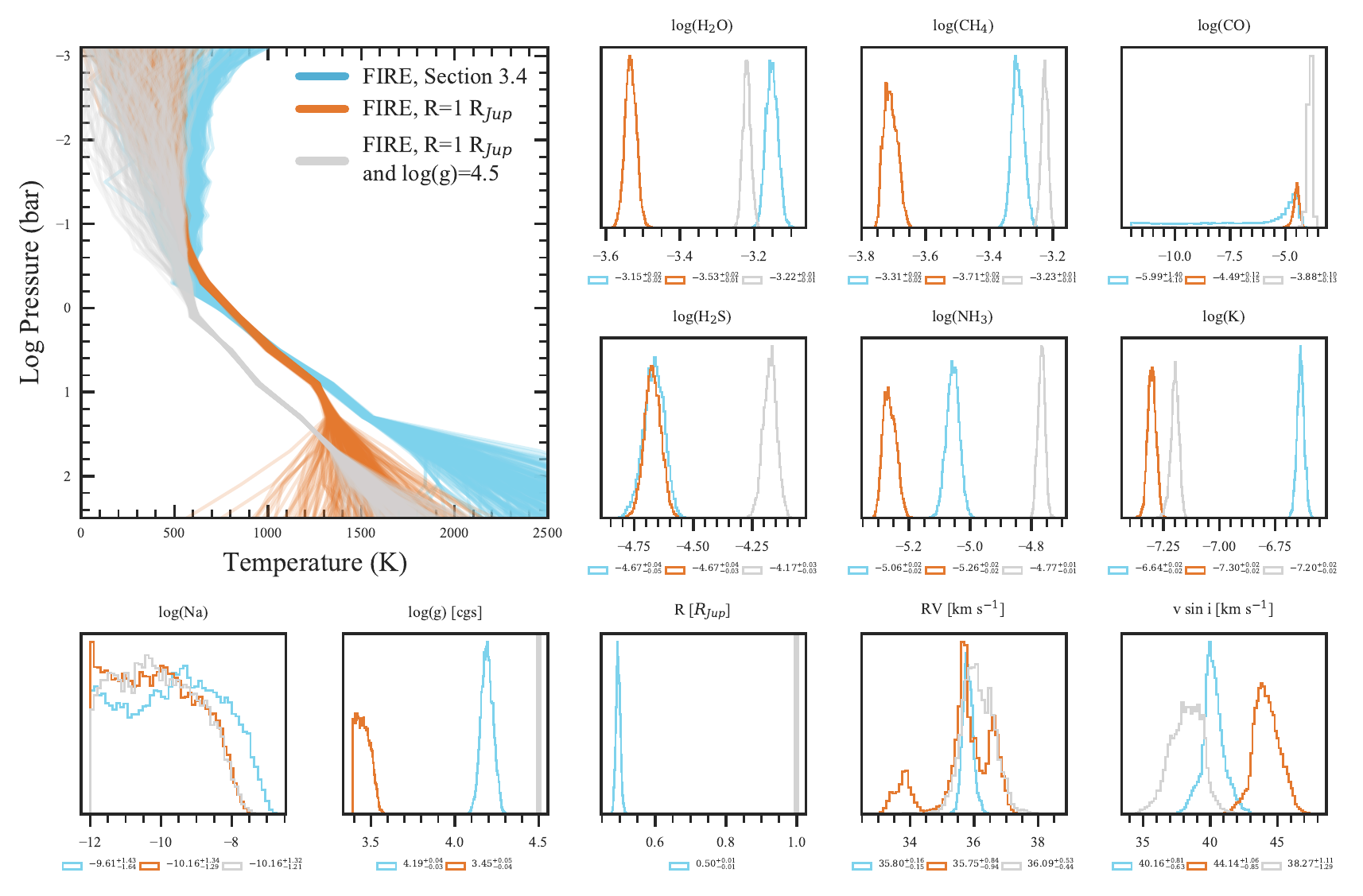}
  		\caption{Effect on the retrieved TP profiles and selected posteriors of fixing the radius to 1 R$_{Jup}$ (orange), and then setting the surface gravity to log(g)=4.5 as well. Almost all parameters are significantly affected.}
  		\label{fig:FIRE_fixRandg}
\end{figure*}

Given the unphysically small retrieved radius for U0722, we explored the impact of fixing the retrieved radius to 1 Jupiter radius on the TP profile and abundance constaints. Figure \ref{fig:FIRE_fixRandg} shows the retrieved TP profiles and selected posteriors with this fixed radius in orange compared to when the radius is allowed to vary for the retrieval described in Section \ref{sec:stitch}. The retrieved TP profiles overlap with the previous results for $\sim$2-4 bars but with a slightly different slope, as they are hotter above and cooler below this region. All abundance posteriors shift, including the disappearance of the long low tail for CO, except for H$_{2}$S which remains relatively unchanged. In particular all abundance shifts are to lower values, perhaps balancing out the cooler temperatures in the deep atmosphere where the flux in \text{y} and \textit{J} bands originates. The radial velocity posterior is strangely trimodal; however, one peak does correspond to the previous value. The $v$ sin $i$ posterior does shift as well. Importantly, the surface gravity decreases to an unphysically small median value of log(g)= 3.45. Figure \ref{fig:CornerPlot} in the Appendix shows the correlations between parameters for the retrieval discussed in Section \ref{sec:stitch}. Gravity is notably positively correlated with the abundances of major absorbers (H$_{2}$O, CH$_{4}$, NH$_{3}$) as well negatively correlated with the radius (through the scaling factor (R/D)$^{2}$). As gravity increases, the column optical depth decreases, leading to more flux particularly in the peak of each band (see Figure 2 of \citealp{Line2015}). Therefore, when the radius is fixed to the larger value of 1 R$_{Jup}$, the retrieval converges on a smaller gravity value to balance out the increase in flux (which is also consistent with the lowered abundances).  Due to this questionably low surface gravity, we do not use this fixed radius retrieval to compare to SpeX and grid models in Section \ref{sec:discandcon}.

\begin{figure}[htbp]
         \centering
        \includegraphics[width=.95\linewidth]{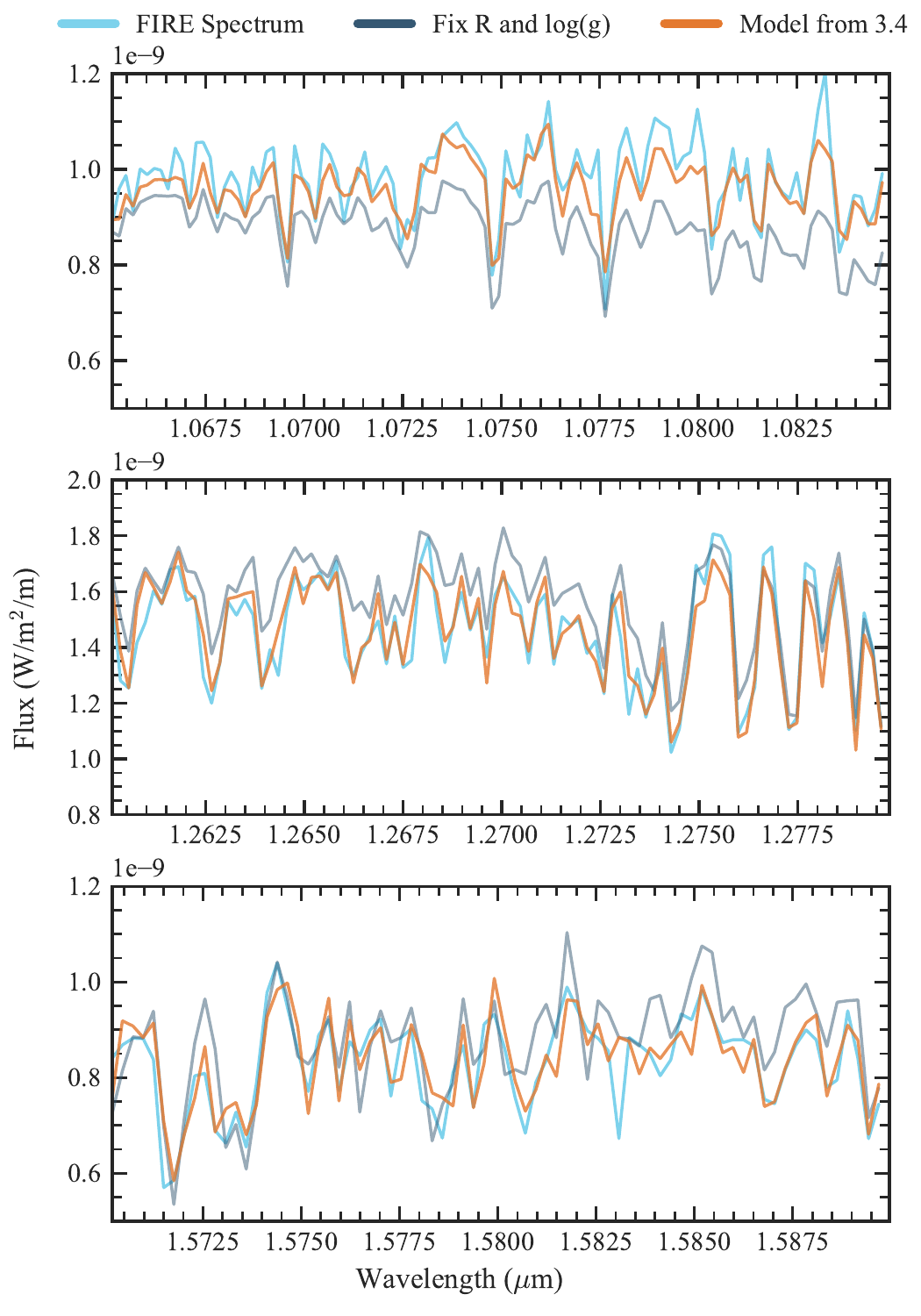}
  		\caption{Select regions of the model spectrum from the median retrieved parameters when the radius and log(g) are fixed to 1 R R$_{Jup}$ and 4.5, respectively, compared to the median model when both are allowed to vary (from Section \ref{sec:stitch}) and the FIRE spectrum of U0722. The fit to the data is worse when the radius and gravity are not allowed to vary in the retrieval.}
  		\label{fig:fixRandg_model}
\end{figure}

We next considered fixing the surface gravity simultaneously with the radius, choosing log(g)=4.5 consistent with a recent grid model study of U0722 \citep{Leggett2021}. The resulting TP profile and posteriors are shown in Figure \ref{fig:FIRE_fixRandg} in grey. Since the gravity can no longer be decreased to account for the increased flux from a 1 R$_{Jup}$ object, the retrieved TP profile is significantly colder than the other two plotted retrievals for the entire region of the atmosphere where it is well-constrained, about 0.5 to 50 bars. We will compare this TP profile to grid model predictions in Section \ref{sec:discandcon}. All abundance posteriors shift significantly compared to when the radius and surface gravity are allowed to vary, though the direction of this shift varies for different species. The radial velocity and $v$ sin $i$ posteriors change slightly as well, though the former does not exhibit the same trimodal behavior as when the radius alone is fixed. While these constraints seem more plausible than those obtained when only the radius was fixed to a certain value, the fit to the data does suffer compared to when the radius and gravity are allowed to vary, as shown in Figure \ref{fig:fixRandg_model}. Furthermore, our choice of fixed radius and surface gravity, while guided by previous studies of this object, are still relatively arbitrarily chosen. As the benefit of a retrieval analysis is to obtain constraints with fewer a priori assumptions than fitting the data to grid models, we do not use this fixed radius and gravity retrieval as our ``preferred" FIRE retrieval in Section \ref{sec:discandcon} for comparing to SpeX and grid model results.

\subsection{Cloudy Retrieval}\label{sec:clouds}
\begin{figure*}[htbp]
         \centering
        \includegraphics[width=.95\linewidth]{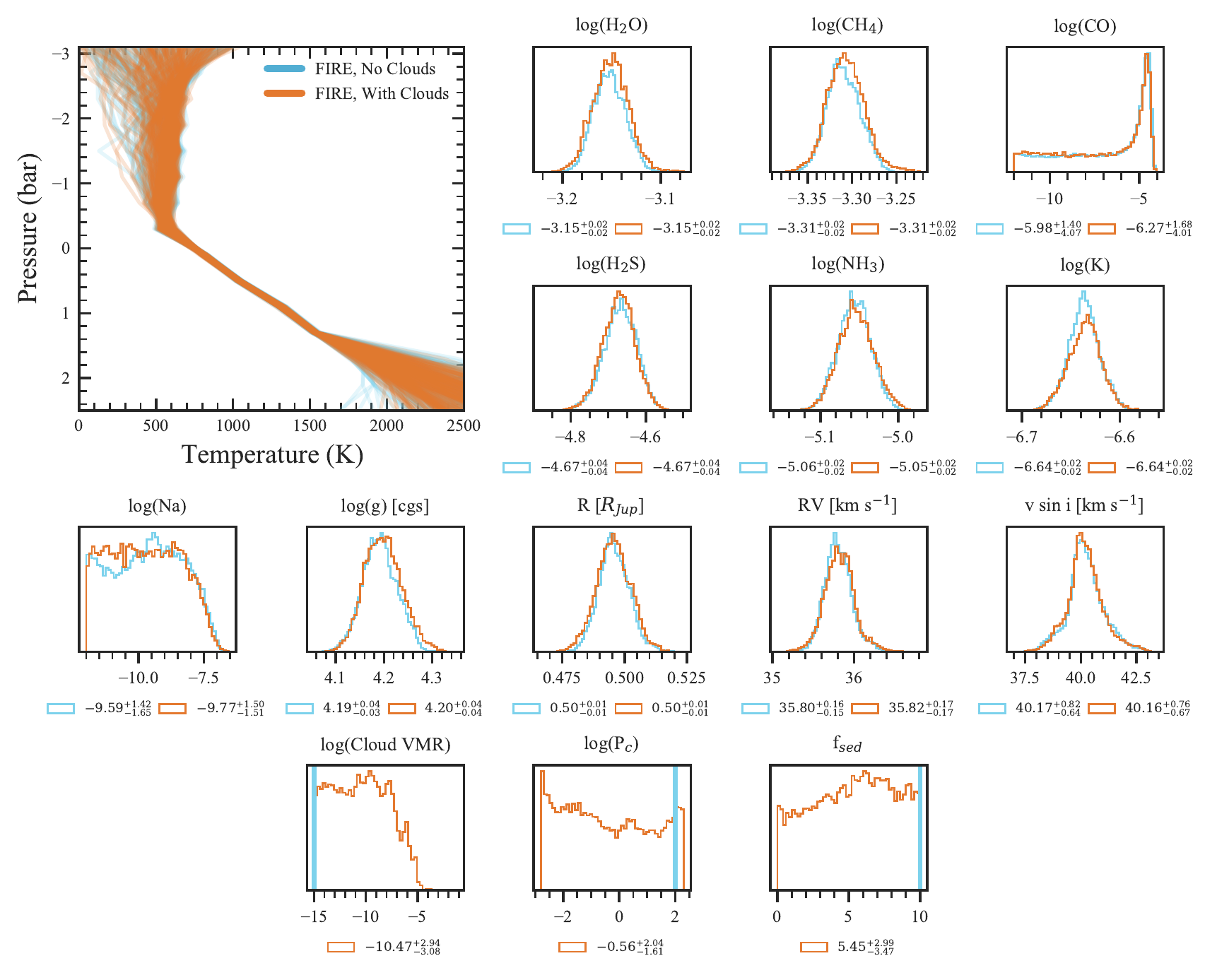}
  		\caption{Effect on the retrieved TP profiles and selected posteriors of allowing the cloud parameters to vary (orange), instead of being fixed to values corresponding to optically thin clouds (blue).  We find little evidence for optically thick clouds in the atmosphere of U0722.}
  		\label{fig:FIRE_cld}
\end{figure*}

In Section \ref{sec:init} we fixed our cloud parameters to be consistent with an optically thin cloud that would not affect the emission spectrum when it appeared our model was perhaps too flexible. After incorporating the preceding changes we found improved our results (fixing $\beta$, changing the wavelength limits to match \citetalias{Bochanski2011}, sampling one data point per resolution element, updating the CH$_{4}$ and NH$_{3}$ opacities, and ignoring regions victim to poor order stitching), we allowed the cloud parameters to vary once again. Figure \ref{fig:FIRE_cld} shows the resulting TP profiles and posteriors in grey. The cloud parameters are unconstrained, except for a potential upper limit on the cloud volume mixing ratio. Other parameters are relatively unaffected, suggesting the inclusion of cloud opacity has little impact. This lack of evidence for optically thick clouds in the atmosphere of U0722 are consistent with expectations for late T dwarfs \citep[e.g.][]{Kirkpatrick2005} as well as other T dwarf retrieval studies at lower spectral resolution \citep[e.g.][]{Line2017, Zalesky2022}.

\section{Discussion and Conclusions} \label{sec:discandcon}
\subsection{SpeX vs FIRE}\label{disc:spexvfire}

\begin{figure*}[htbp]
         \centering
        \includegraphics[width=.95\linewidth]{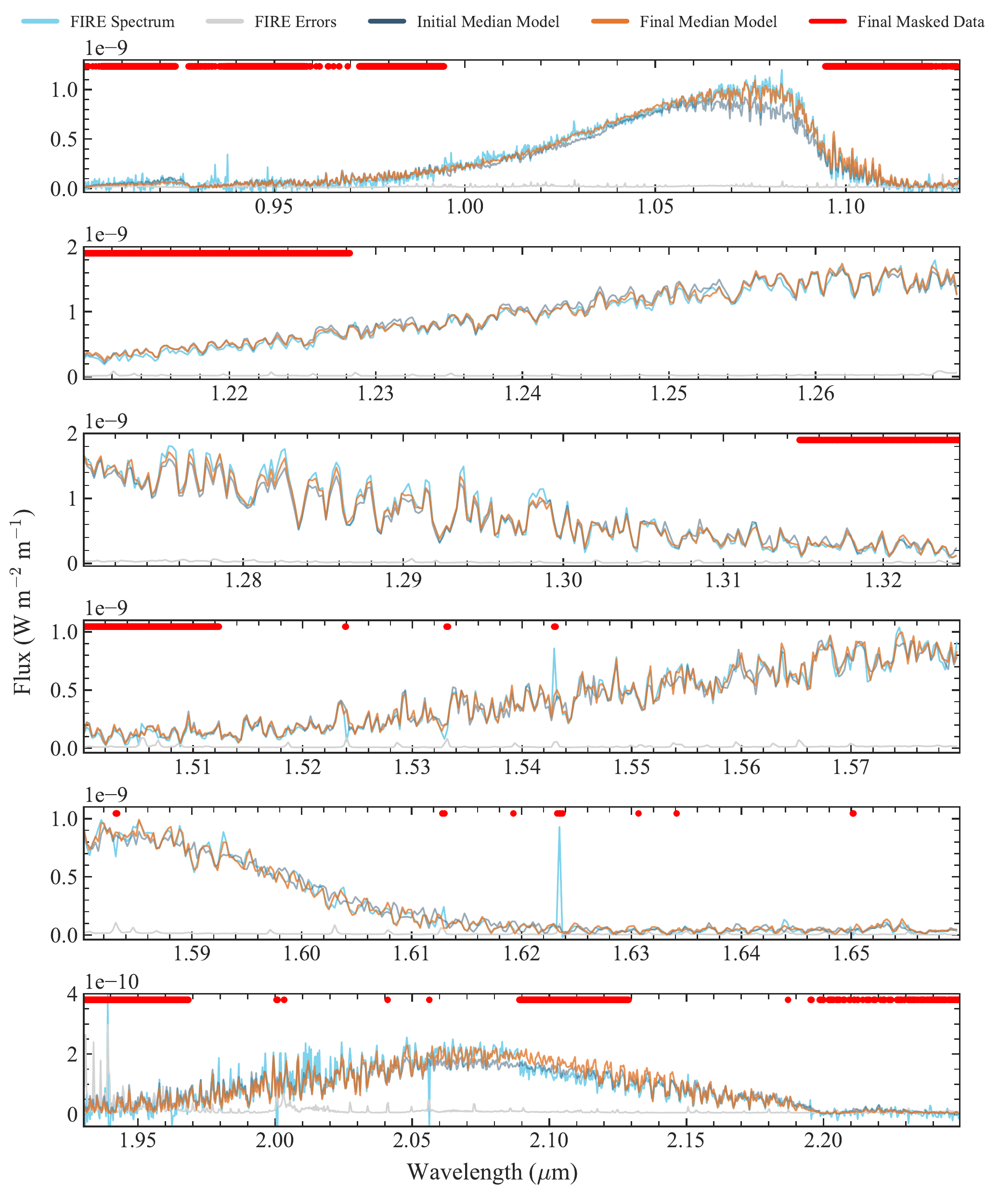}
  		\caption{Comparison of the FIRE spectrum (light blue) with the median model of our initial retrieval (navy) and the median model of our “final” preferred FIRE retrieval from Section \ref{sec:stitch} (orange). The FIRE errors are shown in gray. Red lines at the top of each panel indicate sections of the FIRE spectrum that were removed from our final analysis as described in Sections \ref{sec:init}, \ref{sec:resel}, and \ref{sec:stitch}. Our final preferred median model spectrum does a better job fitting the FIRE data, particularly the peak of \textit{y} band and the CH$_{4}$ features in \textit{H} band.}
  		\label{fig:FIRE_modelsvdata}
\end{figure*}

We choose the FIRE retrieval from Section \ref{sec:stitch}, with pieces of the spectrum affected by order stitching removed and all the preceding changes, as our ``preferred” FIRE retrieval. While Sections 3.5 - 3.8 explore various other tests we performed, the retrieval from Section \ref{sec:stitch} uses the entire wavelength range while providing more physical constraints than those gained when using the new alkali opacities or fixing the radius for example. Table \ref{tab:tests} lists each change to our retrieval framework we tested, the section of the paper where it is discussed, and whether or not it was applied for our preferred retrieval.   Figure \ref{fig:FIRE_modelsvdata} shows the FIRE spectrum compared to the median model spectrum from the initial retrieval described in \ref{sec:init} and the median model spectrum from this final preferred FIRE retrieval. Red lines at the top of each panel indicate data points removed from our analysis due to spurious flux values and high telluric absorption (Section \ref{sec:init}) or suspected problems with order stitching (Section \ref{sec:stitch}). In Figure \ref{fig:FIRE_modelsvdata}, the model from our final retrieval fits the FIRE spectrum much better than for our initial retrieval, particularly the peak of \textit{y} band and the CH$_{4}$ features in \textit{H} band, as discussed in Section \ref{sec:linelist}. However, though the posteriors of the retrieved parameters are quite different, there are large sections of the spectrum where both models reproduce the observed spectrum relatively well, as should be expected for a data-driven model fitting method. Most of the tests of our framework outlined in Section \ref{sec:results}, other than updating the line lists, were motivated by intuition or unphysical results rather than poor fits to the data. As such, we do not plot the median models from all our different retrieval runs in the interest of brevity and clarity.

\begin{figure*}[htbp]
         \centering
        \includegraphics[width=.95\linewidth]{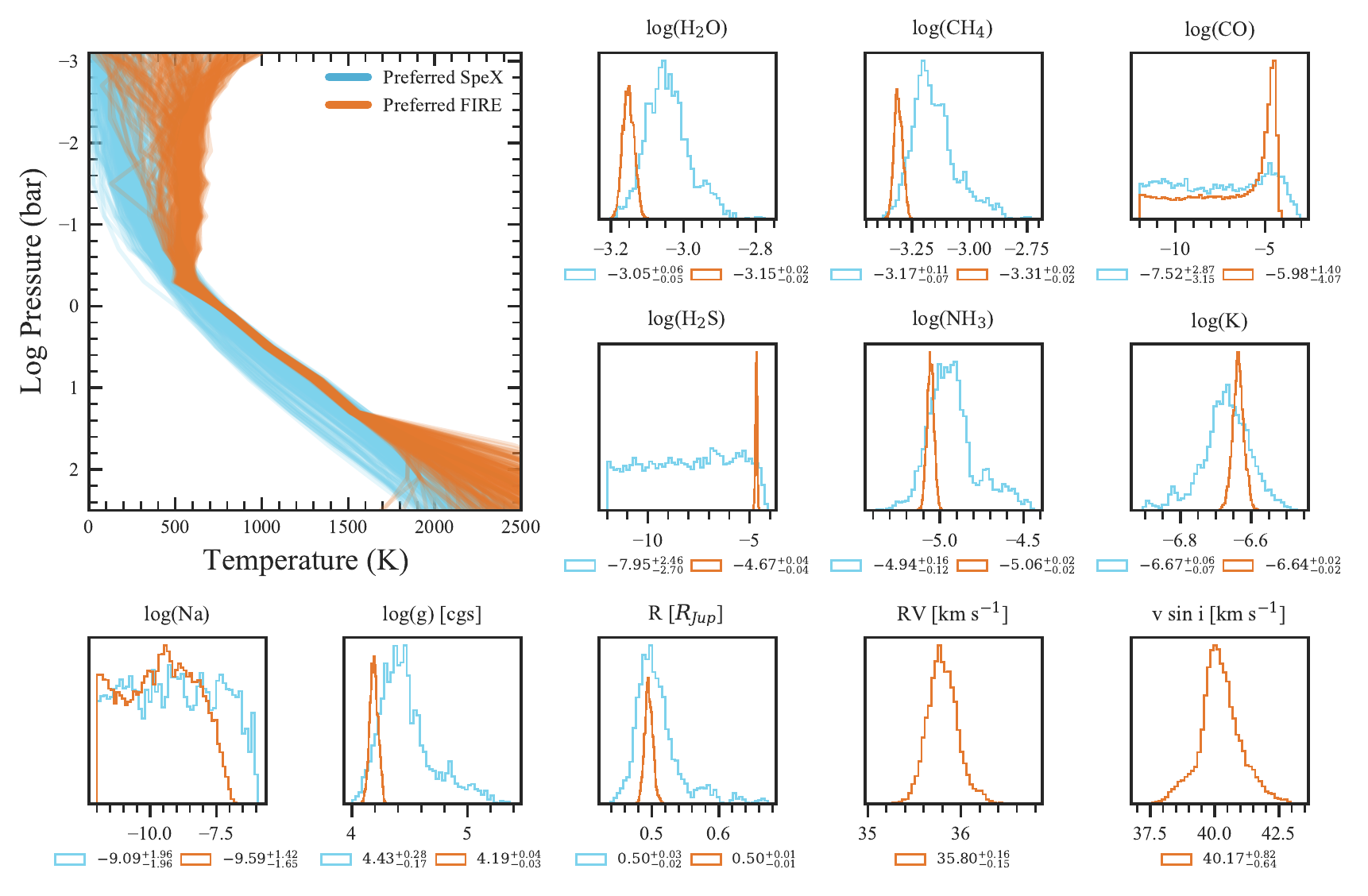}
  		\caption{Retrieved TP profiles and selected posterior distribution for selected parameters for SpeX and FIRE spectra of U0722 after making the changes discussed in Section \ref{sec:results}. }
  		\label{fig:FIREvSpeX_final}
\end{figure*}

Figure \ref{fig:FIREvSpeX_final} shows how much the constraints on our retrieved posteriors improve when using the R $\sim$ 6000 FIRE spectrum compared to the R $\sim$ 100 SpeX spectrum. The ``preferred" SpeX retrieval here uses the same updated line lists as described in Section \ref{sec:linelist}. The H$_{2}$O, CH$_{4}$, and NH$_{3}$ abundance posteriors with FIRE are $\sim$3-6$\times$ more precise than those from SpeX. Similarly, we can constrain the surface gravity to within about 0.04 dex, about 5$\times$ more precise than SpeX. Furthermore, between $\sim$0.5 to 20 bars, the TP profile is constrained within 100 K compared to the much wider 500 K spread for SpeX. Finally, our retrieval on the medium resolution FIRE spectrum allows constraints on parameters such as H$_{2}$S abundance, CO abundance, radial velocity, and $v$ sin $i$ which are not constrained with the SpeX spectrum. However, while the preferred FIRE retrieval yields very precise constraints, they may not be accurate descriptions of the characteristics of U0722 and the uncertainties do not include any systematic effects, which we will discuss further below.

\begin{table*}[htbp]
\centering
\caption{Changes to our retrieval framework tested for application to the FIRE spectrum of U0722.} \label{tab:tests}
\begin{tabular}{lll}
\hline
\hline
Change & Section of Paper & In preferred retrieval? \\
\hline
Add radial velocity and $v$ sin $i$ & \ref{subsec:framework} & yes \\
Mask regions of total telluric absorption & \ref{sec:init} & yes \\
Remove second T-P smoothing hyperparameter & \ref{sec:init} & yes \\
Fix cloud parameters to optically thin values & \ref{sec:init} & yes \\
Limit data to 0.9-2.35 $\mu$m & \ref{sec:resel} & yes \\
Take every 4th data point to not oversample the FIRE resolution element & \ref{sec:resel} & yes \\
Update NH$_{3}$ and CH$_{4}$ opacities & \ref{sec:linelist} & yes \\
Mask data points affected by issues with orders stitching & \ref{sec:stitch} & yes \\
Limit analysis to subsections of the spectrum, \textit{J}-\textit{K} and \textit{H}-\textit{K} & \ref{sec:partret} & no \\
Update alkali opacities & \ref{subsec:alkali} & no \\
Fix radius to 1 R$_{Jup}$ & \ref{sec:radius} & no \\
Fix log(g)=4.5 and R = 1 R$_{Jup}$ & \ref{sec:radius} & no \\
Allow cloud parameters to vary & \ref{sec:clouds} & no \\
\hline
\end{tabular}
\end{table*}

Table \ref{tab:props} lists physical parameters of U0722 for both retrievals, as well as from previous studies which will be discussed below. To calculate a number of these quantities and their uncertainties from the parameters in our retrieval framework, we take 5000 random samples of our posterior. With these precise constraints on the molecular abundances, we can consider the metallicity and C/O ratio as most of the metal content in these cool brown dwarf atmospheres is contained in H$_{2}$O and CH$_{4}$, with some contribution from CO and NH$_{3}$. We calculate these quantities from our retrieved abundances following Equations 1 and 2 from \citet{Zalesky2022} and assuming solar abundances from \citet{Lodders2010}. To account for condensation processes that can deplete atmospheric oxygen, we multiply our H$_{2}$O abundance by 1.3 to better approximate the intrinsic metallicity and C/O as in \citet{Zalesky2022}. Our retrieved abundances yield [M/H] = -0.10$^{+0.02}_{-0.02}$ dex and C/O = 0.54$^{+0.01}_{-0.02}$ from the FIRE retrieval, with $\sim$3-4$\times$ the precision than from the SpeX retrieval. For comparison, \citet{Zalesky2022} on average constrain [M/H] and C/O to within 0.2 dex with 50 T dwarf spectra at R$\sim$100, although the precision varies among the objects. Though we have achieved precise constraints on these bulk properties, \cite{Calamari2022} carried out a retrieval study of GI 229B and showed the 1.3 oxygen scaling factor lead to a calculated C/O that was unexpectedly inconsistent with measurements of the primary star. They suggest better understanding of the potential oxygen sinks in brown dwarf interiors could refine the best way to connect measured atmospheric C/O to the bulk value.

For each sample of our posterior, we also generate a low-resolution spectrum over 0.3 to 250 $\mu$m, which we integrate and use to compute the bolometric luminosity L$_{Bol}$ and effective temperature T$_{eff}$. \citet{Leggett2012} report a luminosity range of log(L$_{Bol}$/L$_{Sun}$) = -6.05-6.17 from observed spectra of U0722 over $\sim$ 0.7 - 4 $\mu$m. The calculated log(L$_{Bol}$/L$_{Sun}$) from the FIRE results of -6.23$^{+0.20}_{-0.09}$ is consistent with this literature value, but the luminosity from our SpeX results log(L$_{Bol}$/L$_{Sun}$) = -6.42$^{+0.08}_{-0.05}$ is significantly lower. We also note that the luminosity constraints from the SpeX retrieval are more precise than those from the FIRE retrieval. While between 0.5 - 20 bars the TP profile is much better constrained by the FIRE retrieval, outside of this range the temperatures vary considerably and are hotter than the corresponding SpeX values, leading to a calculated luminosity that is both higher and less well-constrained. The  effective temperatures calculated from the luminosities and radius values unsurprisingly show the same behavior.

\begin{table*}
\begin{center}
\renewcommand{\arraystretch}{1.25}
\caption{Parameters of U0722 calculated from this work and previous studies.} \label{tab:props}
\begin{tabular}{lllll}
\hline
\hline
Parameter & FIRE Retrieval  & SpeX Retrieval & BT-Settl Model Grid Fit & Tuned ATMO Model Grid Fit \\
 & (this work) & (this work) & \citet{Bochanski2011} & \citet{Leggett2021} \\
\hline
Wavelength Range ($\mu$m) & 0.9 - 2.35 & 0.9 - 2.35 & 0.9 - 2.35 & 0.6 - 5.1 \\
Spectral Resolution & 6000 & 100 & 6000 & 180-500\tablenotemark{f} \\
log(g) (cgs) & 4.19$^{+0.04}_{-0.03}$ & 4.43$^{+0.28}_{-0.17}$ & 4.0$^{+0.3}_{-0.3}$ & 4.5 \\
Radius (R$_{Jup}$) & 0.5$^{+0.01}_{-0.01}$ & 0.5$^{+0.03}_{-0.02}$ & 1.14$^{+0.46}_{-0.33}$ \tablenotemark{a} & 1.12\tablenotemark{g} \\
Distance (pc) & Fixed at 4.12 & Fixed at 4.12 & 4.6 (restricted)\tablenotemark{b} & Fixed at 4.12 \\
 & & & 49 (unrestricted) & \\
Mass (M$_{Jup}$) & 1.5$^{+0.1}_{-0.1}$ & 1.9$^{1.2}_{-0.6}$ & 5.24\tablenotemark{c} & 15\tablenotemark{g} \\
C/O & 0.54$^{+0.01}_{-0.02}$ & 0.55$^{+0.08}_{-0.06}$ & 0.55\tablenotemark{d} & 0.55\tablenotemark{h} \\
{[M/H]} (dex) & -0.10$^{+0.02}_{-0.02}$ & -0.02$^{+0.07}_{-0.06}$ & 0.0\tablenotemark{d} & 0.0\tablenotemark{h} \\
log(L$_{Bol}$/L$_{Sun}$) & -6.23$^{+0.20}_{-0.09}$ & -6.42$^{+0.08}_{-0.05}$ & -6.11 (restricted)\tablenotemark{e} & -5.99\tablenotemark{e} \\
 & & & -5.52 (unrestricted) & \\
T$_{eff}$ (K) & 711$^{+84}_{-35}$ & 638$^{+32}_{-15}$ & 500$^{+50}_{-50}$ (restricted) & 540 \\
  & & & 700$^{+50}_{-50}$ (unrestricted) & \\

\hline
\end{tabular}
\end{center}
\tablenotetext{a}{Calculated from reported log(g) and mass}
\tablenotetext{b}{\citet{Bochanski2011} perform two fits, one with distance as a free parameter and one where the model-derived distance must be within 5$\sigma$ of the parallax measurement from \citet{Lucas2010}.}
\tablenotetext{c}{Determined from evolutionary models of \citet{Baraffe2003} and the best fit T$_{eff}$ and log(g)}
\tablenotetext{d}{Considered models from the BT-Settl model grid \citep{Allard2010} assuming solar metallicity and C/O using solar abundances from \citet{Asplund2009}}
\tablenotetext{e}{Calculated from radius and T$_{eff}$.}
\tablenotetext{f}{Used flux-calibrated SED with datasets from multiple instruments published by \citet{Lucas2010, Leggett2012, Miles2020}}
\tablenotetext{g}{Determined from evolutionary models of \citet{Phillips2020}}
\tablenotetext{h}{Considered models from the ATMO grid \citep{Phillips2020} assuming solar C/O and with [M/H] = -0.5, 0.0, and +0.3 dex using solar abundances reported by \citet{Asplund2009} and updated by \citet{Caffau2011}}
\end{table*}

However, both our SpeX and FIRE retrievals do result in an unphysically small radius of about 0.5 R$_{Jup}$, potentially calling the accuracy of our other constraints into question. The parameter actually constrained by CHIMERA is the radius-to-distance scaling factor (R/D)$^{2}$; however, U0722 has a well-constrained distance from parallax measurements (4.12 $\pm$ 0.04 pc, \citealt{Leggett2012}) that is most likely not the source of our impossibly small radius. From the Sonora Bobcat evolutionary tracks \citep{Marley2021}, the minimum possible radius for even a 10 Gyr object at subsolar metallicity is 0.75 R$_{Jup}$. Furthermore, this small radius combined with our median retrieved $v$ sin$i$ would imply an unreasonably fast rotation period of about 1.55 hours. Figures \ref{fig:SpeXNewXsec} and \ref{fig:Corner_NewXsecs_stitch} show this radius problem particularly gets worse when updating the CH$_{4}$ and NH$_{3}$ cross sections which are necessary to accurately reproduce the line positions seen in the FIRE spectrum. One potential source of error could be the completeness of the \citet{Hargreaves2020} CH$_{4}$ line list in \textit{y} band as discussed in Section \ref{sec:partret}. Furthermore, even with the previously used \citet{Yurchenko2014} line list we were retrieving a smaller radius than expected (0.76 R$_{Jup}$), indicating additional issues that are perhaps only exacerbated by the change to the newer CH$_{4}$ line list.

Further work is needed to assess whether our small radius could be attributed to completeness issues with the CH$_{4}$ line list from \cite{Hargreaves2020}, uncertainties on how to treat alkali opacities, poor photometric calibration, or some other unseen flaw in our modeling framework.  Recently, multiple retrieval studies have found radii smaller than expected from evolutionary models for both L dwarfs \citep{Gonzales2020, Burningham2021, Xuan2022} and T dwarfs \citep{Kitzmann2020, Lueber2022} across different retrieval frameworks and instruments with varying spectral resolutions. Furthermore, \citet{Zhang2021} performed a uniform comparison of 55 late T dwarf spectra with the Sonora Bobcat grid \citep{Marley2021} of forward models using the Bayesian framework Starfish, finding small radii for a number of the studied objects indicating this issue is not solely found in retrieval analyses. Notably, \citet{Zhang2021} fit a SpeX spectrum of U0722, getting T$_{eff}$= 680 $\pm$ 26 K, log(g)=3.6 $\pm$ 0.3 dex, [M/H]= -0.06 $\pm$ 0.2 dex, and R = 0.43 $\pm$ 0.04 R$_{Jup}$, yielding even more unphysical values of the surface gravity and radius than in this work. More retrieval studies for large samples of brown dwarf spectra, particularly including mid-infrared data for the coolest objects and high-quality observations from JWST, will hopefully illuminate the source of this small radius problem. In addition, the growing sample of transiting brown dwarfs from the TESS mission \citep[e.g.][]{Subjak2020, 2022Carmichael} can provide independent tests of the radii predicted by evolutionary models, though their irradiated nature may cause difficulty in making comparisons. 

\subsection{Comparison to Bochanski et al. Results}
We can compare the results from our preferred retrieval on the FIRE spectrum of U0722 to the original analysis of the dataset presented by \citetalias{Bochanski2011}. By comparing the spectrum to line lists, they were able to identify features of H$_{2}$O, CH$_{4}$, and NH$_{3}$, as well as broad absorption from K. Similarly, we are able to place constraints on the abundances of all of these species to within $\pm$ 0.02 dex, indicating the impact of these species on the spectrum even if the retrieved abundances are not exactly accurate. In addition, we are also able to constrain the H$_{2}$S abundance due to a few distinct H$_{2}$S features in the spectrum (see Section \ref{subsec:medreslinelist}) unidentified in the previous study. 

\citetalias{Bochanski2011} fit the FIRE spectrum of U0722 with BT-Settl models of \citet{Allard2011} based on the fitting procedure outlined by \citet{Cushing2008}. The large differences we see in retrieved posteriors when looking at subsections of the spectrum in Section 3.5 reflect similar variance across the model fits of \citetalias{Bochanski2011} of data in different bandpasses. However, the authors find significant differences between the best fitting models and the data in many places across the spectrum. Furthermore,  there are notable discrepancies in our retrieved physical values from those of \citetalias{Bochanski2011} when considering their full spectrum fit (most analogous to our results) as shown in Table \ref{tab:props}. While their log(g) 4.0$^{+0.3}_{-0.3}$ is consistent, they report a larger mass of 5.24 M$_{Jup}$ from evolutionary models, indicating a much larger radius value of 1.14$^{+0.46}_{-0.33}$ R$_{Jup}$. Due to the large uncertainty on U0722's parallax at the time, \citetalias{Bochanski2011} allow distance instead of the radius to vary when fitting the (R/d)$^{2}$ scaling factor. When allowing this distance to freely vary, their best fit to the entire spectrum also prefers a hotter object with effective temperature 700 $\pm$ 50 K, consistent with the calcualted T$_{eff}$ from our FIRE retrieval but with a far larger distance than the literature value (49 pc vs. 4.12 pc). Our retrieval functionally achieves the same effect with a small (R/d)$^{2}$ scaling factor instead attributed to a small radius given the reliable parallax measurement. However, when the distance is required to be within 5$\sigma$ of the parallax measurement from \citet{Lucas2010}, the authors get a much colder T$_{eff}$ of 500 $\pm$ 50 K. 

Finally, while their measured $v$ sin$i$ of 40 $\pm$ 10 km s$^{-1}$ agrees with our results, their radial velocity measurement of 46.9 $\pm$ 2.5 km s$^{-1}$ is significantly higher than our result of 35.80 $\pm$ 0.15 km s$^{-1}$. Given that the radial velocity was measured in part with models using significantly older line lists for basically all molecules present in this object’s spectrum, such an inconsistency is perhaps expected. Future work to calculate if an updated RV measurement would change the original determination by \citetalias{Bochanski2011} of a thin disk Galactic orbit could provide interesting context to U0722’s potential age and evolutionary trajectory, though the change in radial velocity may not be enough to matter. We use the tool Bayesian Analysis for Nearby Young AssociatioNs $\Sigma$ \citep[BANYAN$\Sigma$,][]{Gagne2018} to determine the probability of U722 being a member of any well-characterized young associations within 150 pc, inputting the previously measured proper motion \citep{Faherty2009} and parallax \citep{Faherty2012} as well as the radial velocity measured in this work.  BANYAN$\Sigma$ gives a 99.9\% likelihood that U0722 is a field brown dwarf. Furthermore, for both velocity parameters our reported precision from \texttt{emcee} is significantly higher than those reported for the measurements by \citetalias{Bochanski2011}. Both the radial velocity and $v$ sin $i$ were calculated by cross-correlating the 1.27 - 1.31 $\mu$m section of FIRE spectrum with a template, either a grid model spectrum or another observation of a T dwarf, which may or may not be a good fit to the data in question. In contrast, the log-likelihood based MCMC approach reported here both considers significantly more data points across the whole spectrum and has been shown to produce smaller error bars than cross-correlation methods \citep{Brogi2019}.

\subsection{Comparison to Grid Models}

\begin{figure}[htbp]
         \centering
        \includegraphics[width=.95\linewidth]{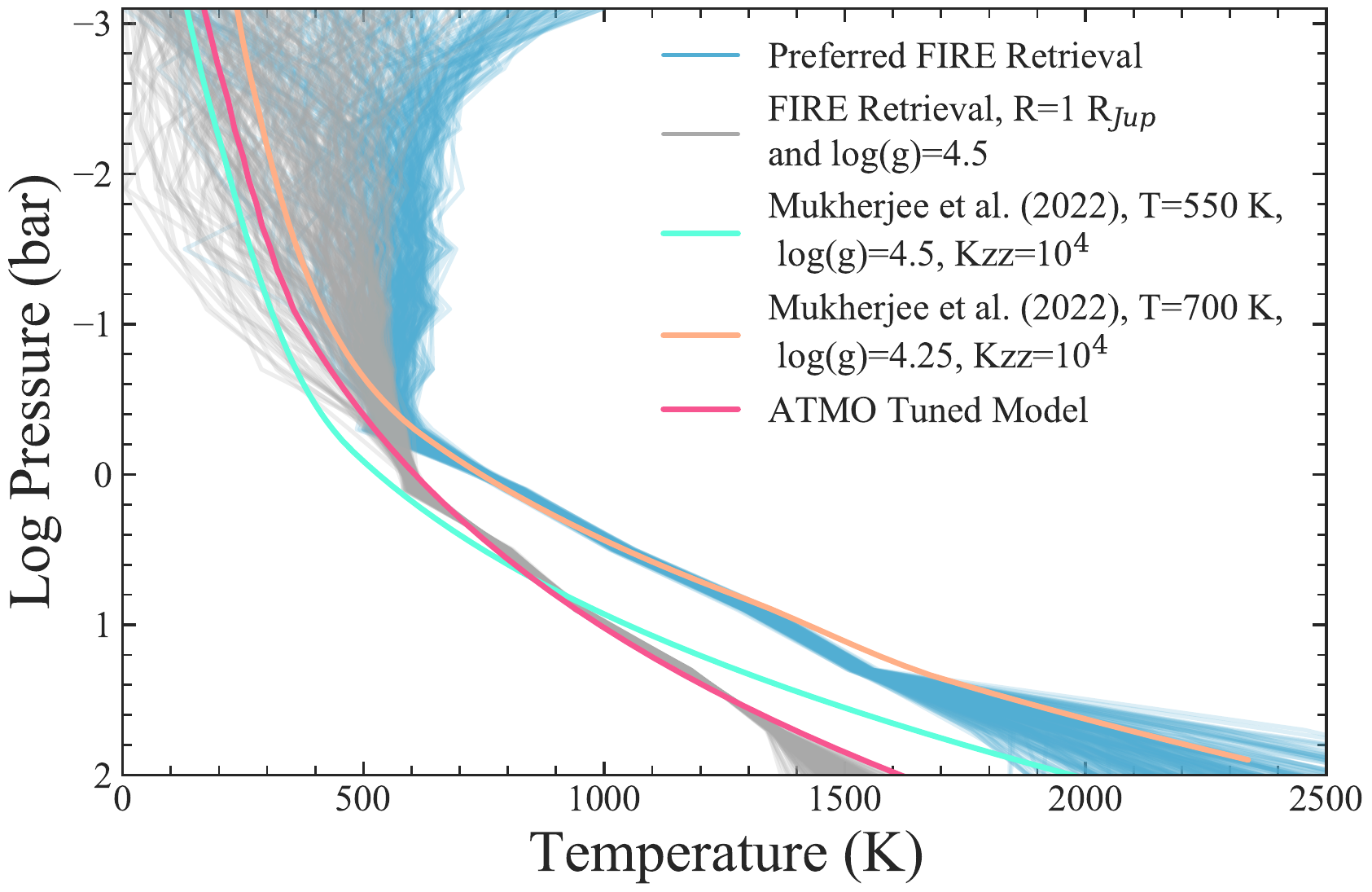}
  		\caption{TP profiles from our preferred FIRE retrieval (blue) and the fixed radius and surface gravity retrieval (grey) to the tuned ATMO model (pink) and two disequilibrium forward models from \citet{Mukherjee2022}. The tuned ATMO model is significantly colder than the retrieved TP profiles from our preferred FIRE retrieval, and has a different gradient in accordance with the adjusted adiabat discussed by \citet{Leggett2021}. However, the retrieved TP profiles when fixing the radius and surface gravity to set values is in excellent agreement with the tuned ATMO model. }
  		\label{fig:PT_gridcomp}
\end{figure}

\begin{figure}[htbp]
         \centering
        \includegraphics[width=.95\linewidth]{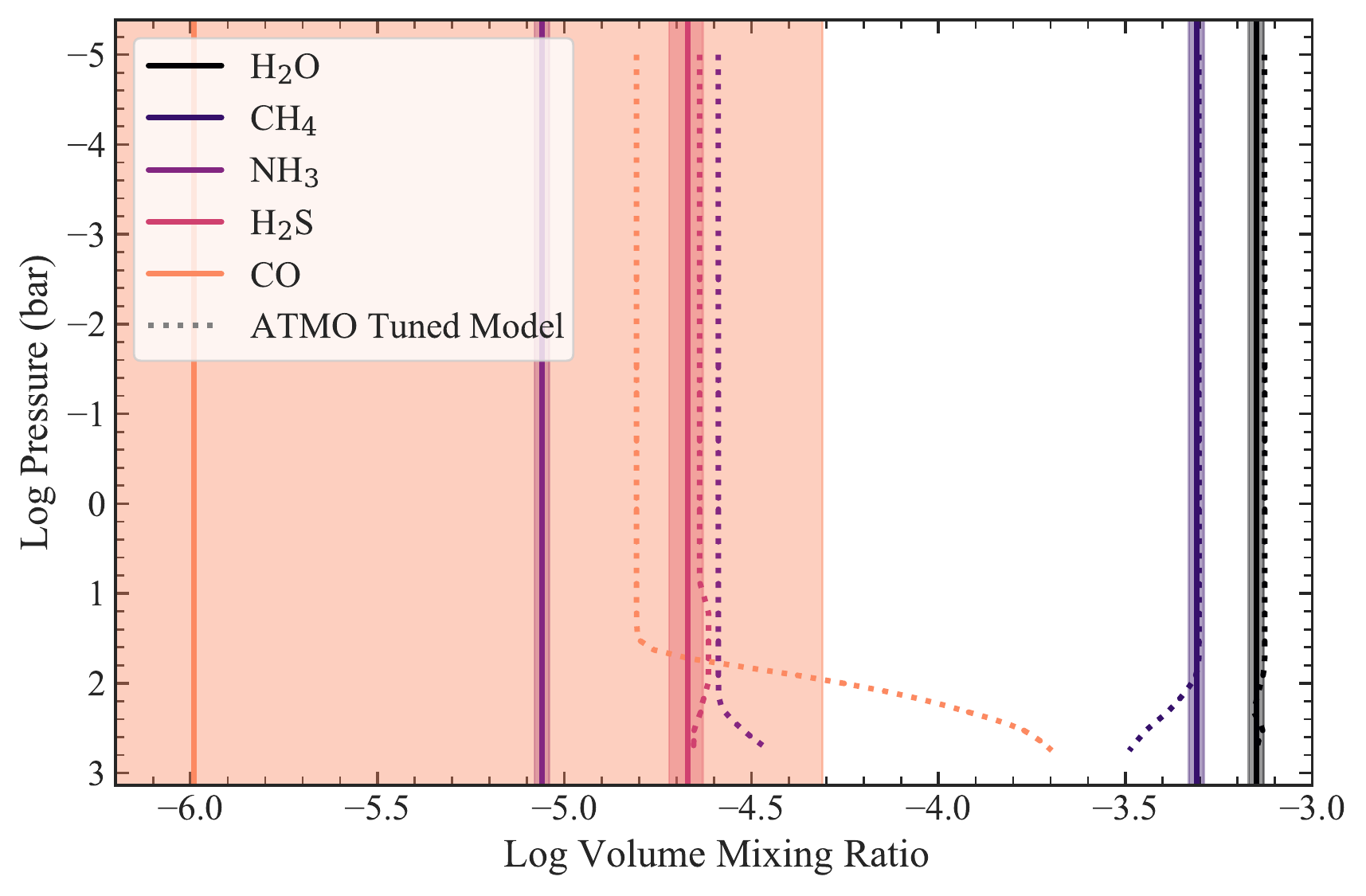}
  		\caption{Molecular volume mixing ratios from the tuned ATMO model (dotted lines) as a function of atmospheric pressure compared to the median and 1$\sigma$ posteriors from our FIRE retrieval (solid lines and shaded regions). With the exception of NH$_{3}$, the abundances are quite similar. }
  		\label{fig:abunds_gridcomp}
\end{figure}

\begin{figure*}[htbp]
         \centering
        \includegraphics[width=.95\linewidth]{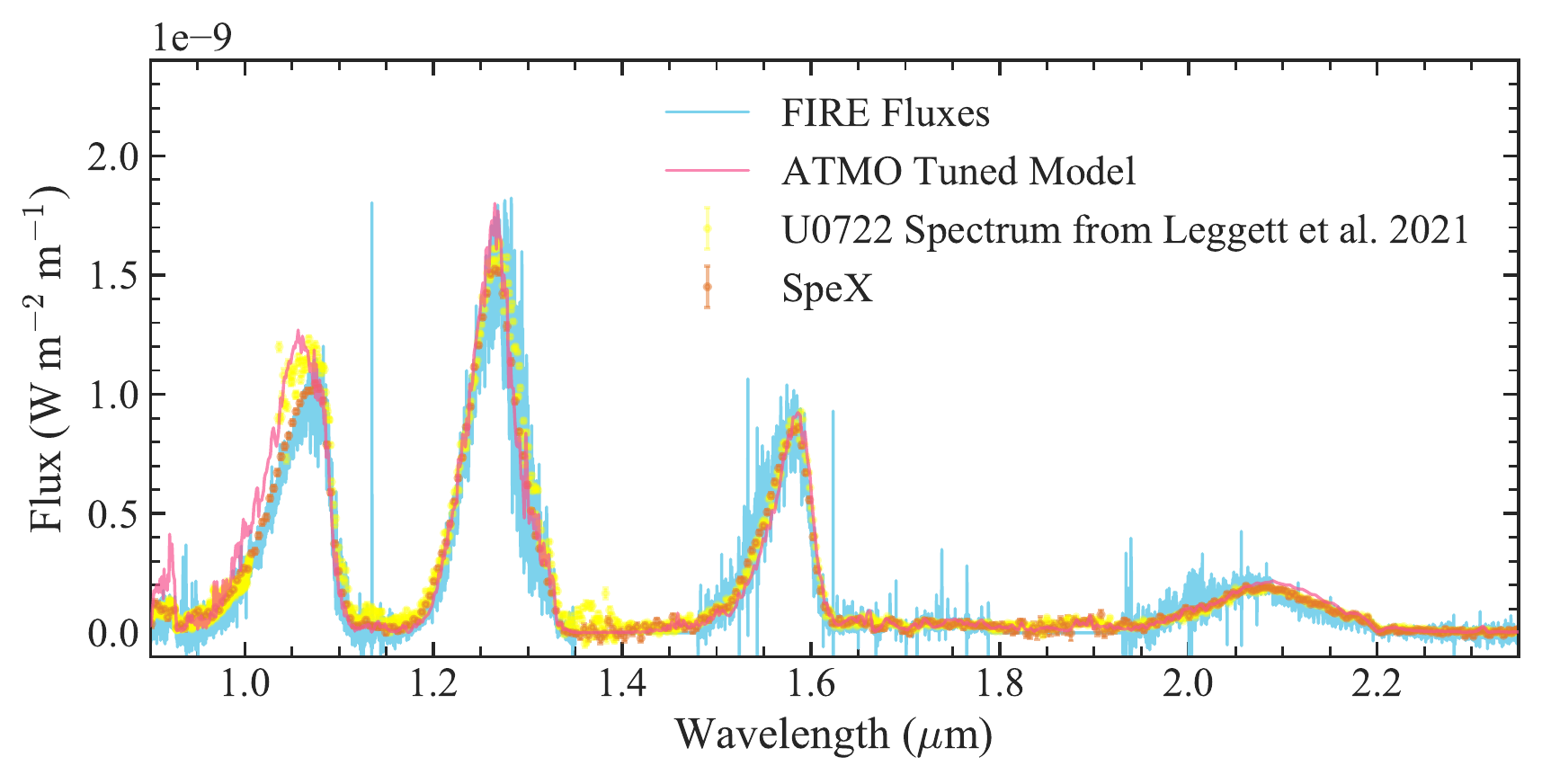}
  		\caption{Comparison of the FIRE (blue) and SpeX (orange) spectra of U0722 analyzed in this work to the spectrum (yellow) and tuned ATMO model (pink) used in \citet{Leggett2021}. There are notable discrepancies in the peak of the \textit{y} band. }
  		\label{fig:Data_ATMOcomp}
\end{figure*}

Given the many advances in line lists over the past decade, it is instructive to look at more recent comparisons of U0722 spectra with grid models. \citet{Miles2020} find that the low-resolution (R $\sim$ 300) \textit{M} band spectrum of U0722 indicates disequilibrium amounts of CO from vertical mixing, consistent with our tentative detection of CO in the FIRE spectrum. \citet{Leggett2021} compare the flux-calibrated 0.6-5.1 $\mu$m spectrum of U0722 created from multiple published observations \citep{Lucas2010,Leggett2012, Miles2020} with the ATMO 2020 grid of models, finding that the models with and without disequilibrium chemistry still have trouble fitting the data well. The authors improve this fit by “tuning” the gradient of the TP profile away from the standard adiabat. As such, we compare our retrieved results to this adiabat-adjusted model in Table \ref{tab:props}, which has an effective temperature of 540 K, log surface gravity (cgs) of 4.5, solar metallicity, log Kzz (cm$^{2}$ s$^{-1}$)=7, and an adiabatic index of 1.27. 

Figure \ref{fig:PT_gridcomp} shows our retrieved TP profiles from the FIRE spectrum of U0722, compared to the abundances of the tuned ATMO model (Tremblin 2021, private communication), as well as two forward grid models with disequilibrium chemistry from \citet{Mukherjee2022}. The tuned ATMO TP profile has a steeper gradient than either of the other grid models, reflecting the adjusted adiabat. Our retrieved TP profiles from the preferred FIRE retrieval are significantly hotter than that of the tuned ATMO model, more closely resembling that of a hotter 700 K object from the \citet{Mukherjee2022} grid, consistent with our calculated T$_{eff}$. The difference in temperature is consistent with the differences in radii- the tuned ATMO model assumed a radius of 1.12 R$_{Jup}$ from the evolutionary models of \citet{Phillips2020}, more than double that of our retrieved value. Thus, a cooler object would be necessary to achieve a similar amount of flux. In fact, when we fixed the radius to 1 R$_{Jup}$ and surface gravity to log(g)=4.5, as discussed in Section \ref{sec:radius}, the retrieved TP profiles are consistent with the tuned ATMO model as shown by the grey curves in Figure \ref{fig:PT_gridcomp}. This agreement is reasonable, as \cite{Leggett2021} similarly set R and log(g) to values from evolutionary model before tuning the TP profile to match the observed data. 

Figure \ref{fig:abunds_gridcomp} compares our retrieved molecular abundances, shown by the solid lines and shaded 1$\sigma$ regions, to the mixing ratios in the tuned ATMO model as a function of pressure in the atmosphere. Unlike the tuned ATMO model which assumes disequilibrium chemistry governed by the eddy diffusion parameter log(K$_{zz}$), our volume mixing ratios for each species are allowed to vary independently and are not tied to one another by any assumptions or parameterizations of the chemical timescales involved. Our retrieved values for CO, CH$_{4}$, and H$_{2}$S are all in agreement with the ATMO abundances above 100 bars, and our H$_{2}$O value posterior is only 0.003 dex away from the ATMO model value. However, our NH$_{3}$ posterior is $\sim$0.5 dex lower, perhaps reflecting the differences in effective temperature and surface gravity between our retrieval results and the tuned ATMO model. Given a measured CO abundance, we can estimate log(K$_{zz}$) by comparing to chemical equilibrium abundances from the Sonora Bobcat structure models \citep{Marley2021} of similar T$_{eff}$ and log(g) as described by \cite{Miles2020}. Since our CO abundance has large uncertainties, our results could imply no vertical mixing for the lowest amount of CO up to log(K$_{zz}$) $\sim$ 3.3 cm$^{2}$ s$^{-1}$ for the maximum amount of CO. The comparatively larger log(K$_{zz}$) = 7  cm$^{2}$ $s^{-1}$ from the tuned ATMO model is needed to get a similar CO abundance for a much colder object with higher surface gravity. While our retrieved surface gravity and radius are unphysical, our molecular abundances are plausible and mostly consistent with grid model predictions. The similarity in molecular abundances is also reflected in the consistency of the C/O ratios and metallicities reported in Table \ref{tab:props}. 

\begin{figure}[htbp]
         \centering
        \includegraphics[width=.95\linewidth]{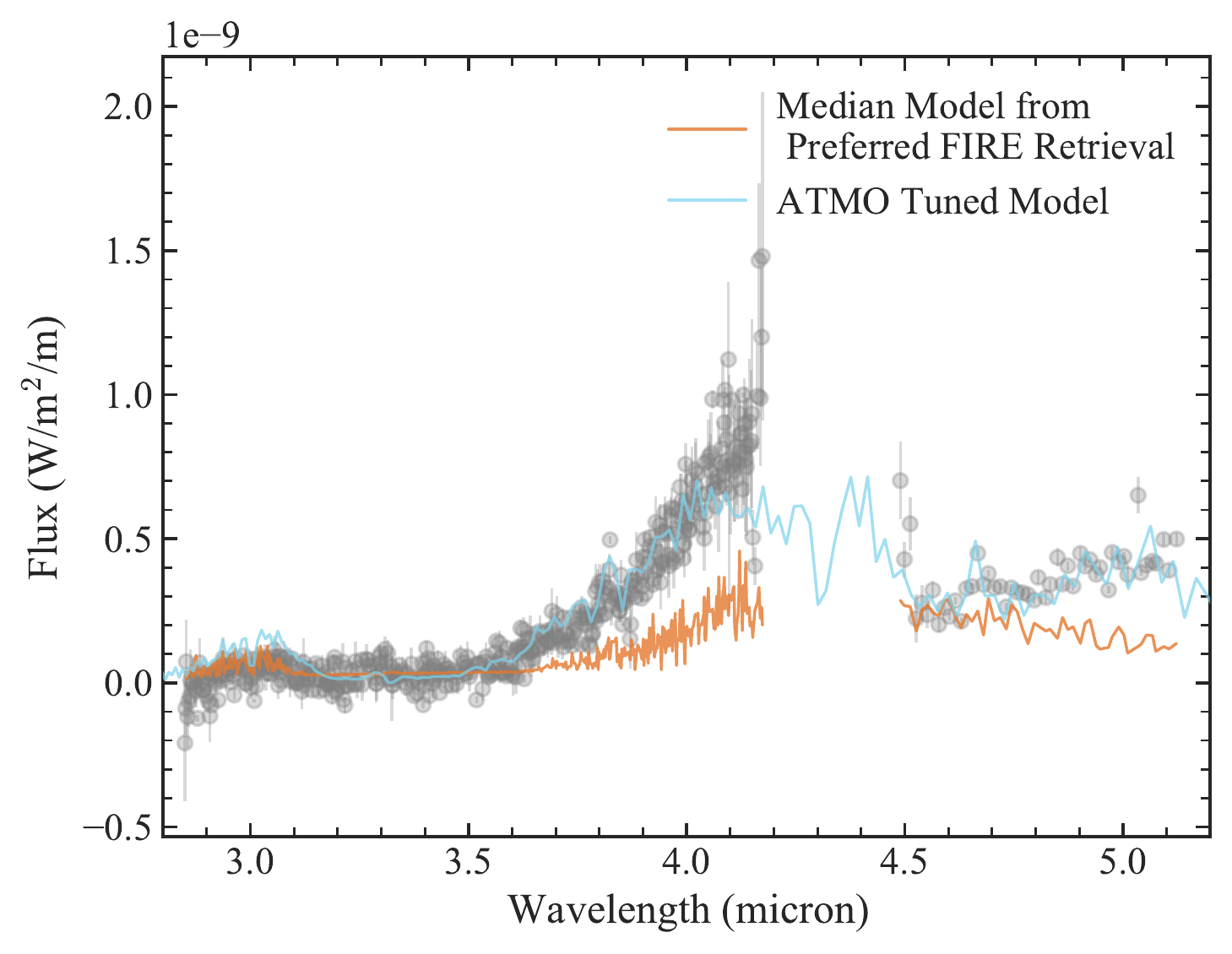}
  		\caption{Comparison of the median model of our preferred FIRE retrieval and the ATMO tuned model to the \textit{L} and \textit{M} band spectra included in the analysis of \cite{Leggett2021}. The FIRE retrieval clearly does not reproduce the measured flux of U0722 at longer wavelengths, whereas the ATMO tuned model gets closer at matching the observed slopes of these regions of the spectrum.}
  		\label{fig:LongWLComp}
\end{figure}

Differences in the data considered perhaps help explain why our retrieval prefers a hotter, smaller object than the tuned ATMO model when the retrieval is solely driven by the data (rather than any assumed radius or gravity value). We note that the NIR portion of the U0722 spectrum used in \citet{Leggett2021} is slightly inconsistent with that of the FIRE and SpeX spectra examined in this work. Figure \ref{fig:Data_ATMOcomp} shows all three datasets, as well as the tuned ATMO model; the peak \textit{y} band flux in particular of the \citet{Leggett2021} is offset in both strength and wavelength. While the source of this discrepancy is unclear, it undoubtedly contributes to differences in our fitted or retrieved values for U0722. However, a more important distinction is the inclusion of longer wavelength data for U0722 in the analysis of \cite{Leggett2021}. 

\begin{figure}[htbp]
         \centering
        \includegraphics[width=.95\linewidth]{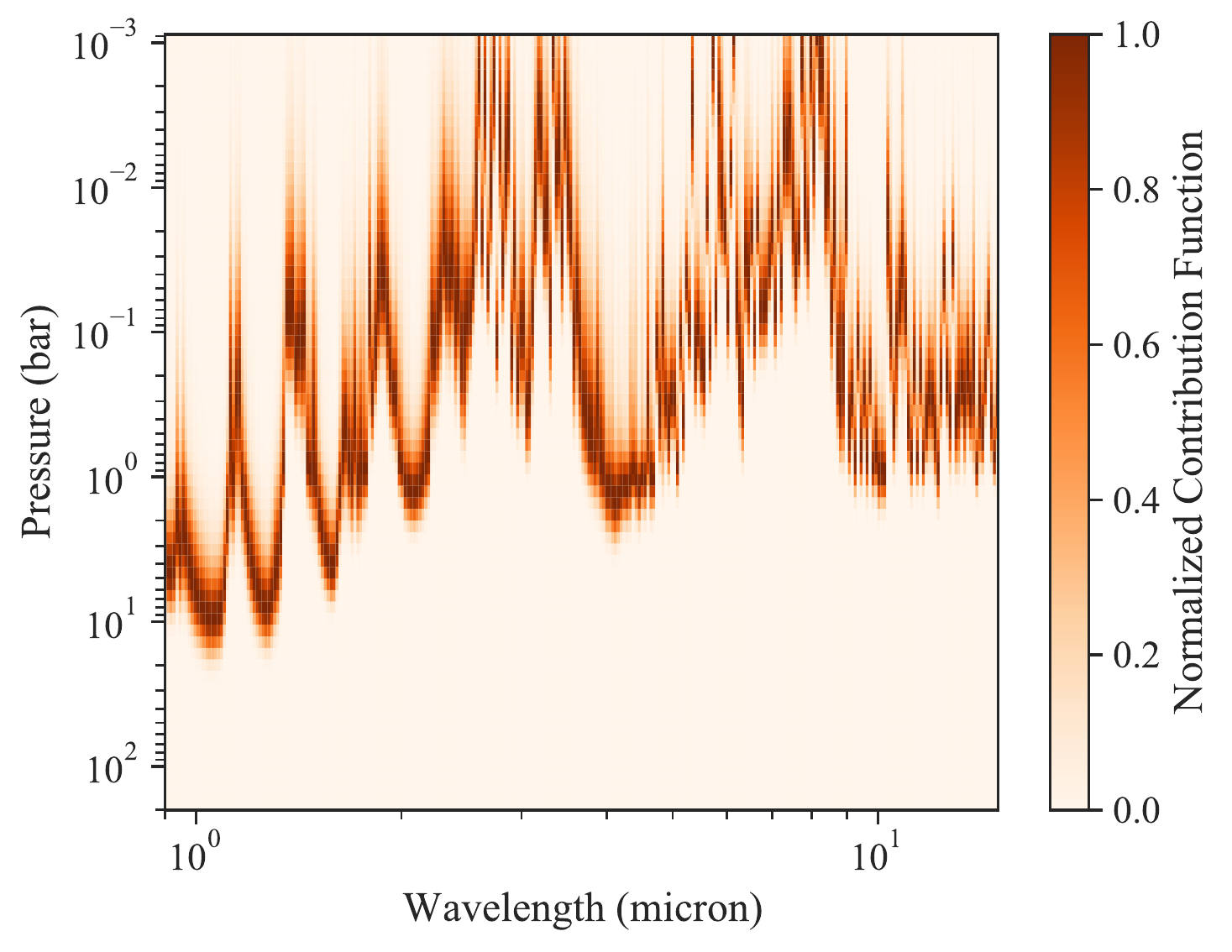}
  		\caption{Normalized contribution functions for a model generated from the median results of preferred FIRE retrieval (which covered 0.9-2.5 $\mu$m) extrapolated out to longer wavelengths.}
  		\label{fig:ContributionFunc}
\end{figure}

The authors included L and M band spectra from \cite{Leggett2012} and \cite{Miles2020}, respectively. Figure \ref{fig:LongWLComp} shows these datasets compared to the median model from our preferred FIRE retrieval extrapolated out to these wavelengths, as well as the tuned ATMO model. Our model clearly underpredicts the observed flux of U0722 and does not reproduce the observed slope in the \textit{M} band data. Figure \ref{fig:ContributionFunc} shows the normalized contribution functions for a model using the median parameter values from our preferred FIRE retrieval, generated at lower resolution and out to $\sim$ 15 $\mu$m. The L and M bands probe the upper atmosphere, and we note that if we were to perform a retrieval on these datasets we would need to extend the atmosphere in our radiative transfer model to pressures below 10$^{-3}$ bars, the current top of the atmosphere in this framework. Notably, the TP profile from our preferred FIRE retrieval is not well constrained at these low pressures in our retrieval, perhaps contributing to the model’s failure to reproduce these data points. Furthermore, the discrepancy in \textit{M} band could be due to less CO present in the median model than is needed to fit the data, as it is a major opacity source in this wavelength region. Figure \ref{fig:abunds_gridcomp} shows that the median CO abundance from our preferred FIRE retrieval is more than a dex less than the amount of CO in the tuned ATMO model. In combination with the unphysically small radius, the inability of our preferred FIRE retrieval to produce fluxes in accordance with longer wavelength observations points to our precise constraints on the atmosphere of U0722 being inaccurate and not necessarily representative of the true physical properties of this object. Future work on combining medium-resolution spectra with broader wavelength coverage data at lower resolution may provide further insight into the best way to obtain more plausible results from an atmospheric retrieval analysis.

\section{Conclusions}\label{sec:conclusion}

In this work, we have applied the CHIMERA atmospheric retrieval framework to a high signal-to-noise, medium-resolution (R $\sim$ 6000) FIRE spectrum of a T9 dwarf.  Key takeaways from this work are as follows:

\begin{enumerate}
    \item \textit{Limitations of the dataset must be taken into account}: In Section \ref{sec:resel}, we show that ensuring only one data point per resolution element is sampled as well as cutting out very noisy regions of data at either end of the spectrum can affect our retrieved posteriors. More dramatically, areas of the spectrum where orders were stitched poorly can negatively bias the retrieval analysis as discussed in Section \ref{sec:stitch}. Future improvements to order stitching methods, or new ways to account for potential stitching issues within a retrieval framework, could help alleviate this problem in future analysis of data from echelle spectrographs. 
    \item \textit{Using different opacity sources may lead to very different results}: Updating the line lists for CH$_{4}$ and NH$_{3}$ to those from \citet{Hargreaves2020} and \citet{Coles2019}, respectively, greatly improved the ability of the forward models to match the line positions in the data, as shown in Figure \ref{fig:CH4ModelComp}. Almost all retrieved posteriors were affected by this change, even for the retrieval on the R $\sim$ 100 SpeX spectrum as shown in Figures \ref{fig:SpeXNewXsec} and \ref{fig:Corner_NewXsecs_stitch}. For example for the FIRE retrieval, changing the line lists lead the median abundances of H$_{2}$O and CH$_{4}$ to increase by $\sim$ 0.3 dex and the surface gravity log(g) to increase by $\sim$0.4 dex. However, the updated line lists also resulted in a radius decreased by about 30\% to an unphysical 0.5 R$_{Jup}$. We also tested different treatments of Na and K opacities in Section \ref{subsec:alkali}, again finding largely disparate results depending on which cross sections were used. While we do recommend the \citet{Hargreaves2020} CH$_{4}$ line list due to its ability to match the line positions in our FIRE spectrum, more comparison with other line lists with regards to completeness as well as treatment of the alkali wings are needed to fully utilize this kind of high-quality data. 
    \item \textit{Medium-resolution retrievals offer very precise constraints, but they may not be accurate}: As shown in Figure \ref{fig:FIREvSpeX_final}, the constraints on the temperature-pressure profile and abundances from our FIRE retrieval of U0722 are significantly more precise than those from the SpeX spectrum. In particular, we are able to retrieve the abundances of H$_{2}$S and tentatively CO which is not possible with the lower resolution spectrum. However, while we do get these precise, stellar-like constraints on atmospheric abundances ($\sim$0.02 dex), the radius is far too small to be physically plausible. This small size is in accordance with our retrieved TP profile being hotter than previous analyses of this object, yielding a similar overall observed flux. Thus, this study joins a growing number of modeling analyses of brown dwarf spectra that have yielded smaller radii than allowed by our understanding of brown dwarf evolution. Furthermore, extrapolating a median model from our retrieval cannot to \textit{L} and \textit{M} band cannot reproduce observations of U0722 at these wavelengths, indicating the constraints from the medium-resolution FIRE spectrum alone do not accurately describe the conditions of this object.
\end{enumerate}

This work is a first foray into the challenges and benefits of applying atmospheric retrieval tools to medium-resolution spectra of brown dwarfs. With the launch of JWST and future ground-based studies, more work is needed to assess how we can improve our current modeling frameworks to address these challenges and unlock the potential for trustworthy, precise constraints on substellar atmospheres from retrievals of medium-resolution spectra.

\begin{acknowledgments}
We thank the anonymous reviewer for their thoughtful comments and suggestions that have improved the quality of this work. We acknowledge use of the lux supercomputer at UC Santa Cruz, funded by NSF MRI grant AST 1828315.  This work benefited from the 2022 Exoplanet Summer Program in the Other Worlds Laboratory (OWL) at the University of California, Santa Cruz, a program funded by the Heising-Simons Foundation. MRL acknowledges support from NASA XRP grant 80NSSC19K0293. JJF acknowledges support from NSF Award Number:1909776 and NASA Award Number: 80NSSC22K0142 as well as the Simons Foundation.
\end{acknowledgments}

\appendix 
\begin{figure*}[h]
         \centering
        \includegraphics[width=.95\linewidth]{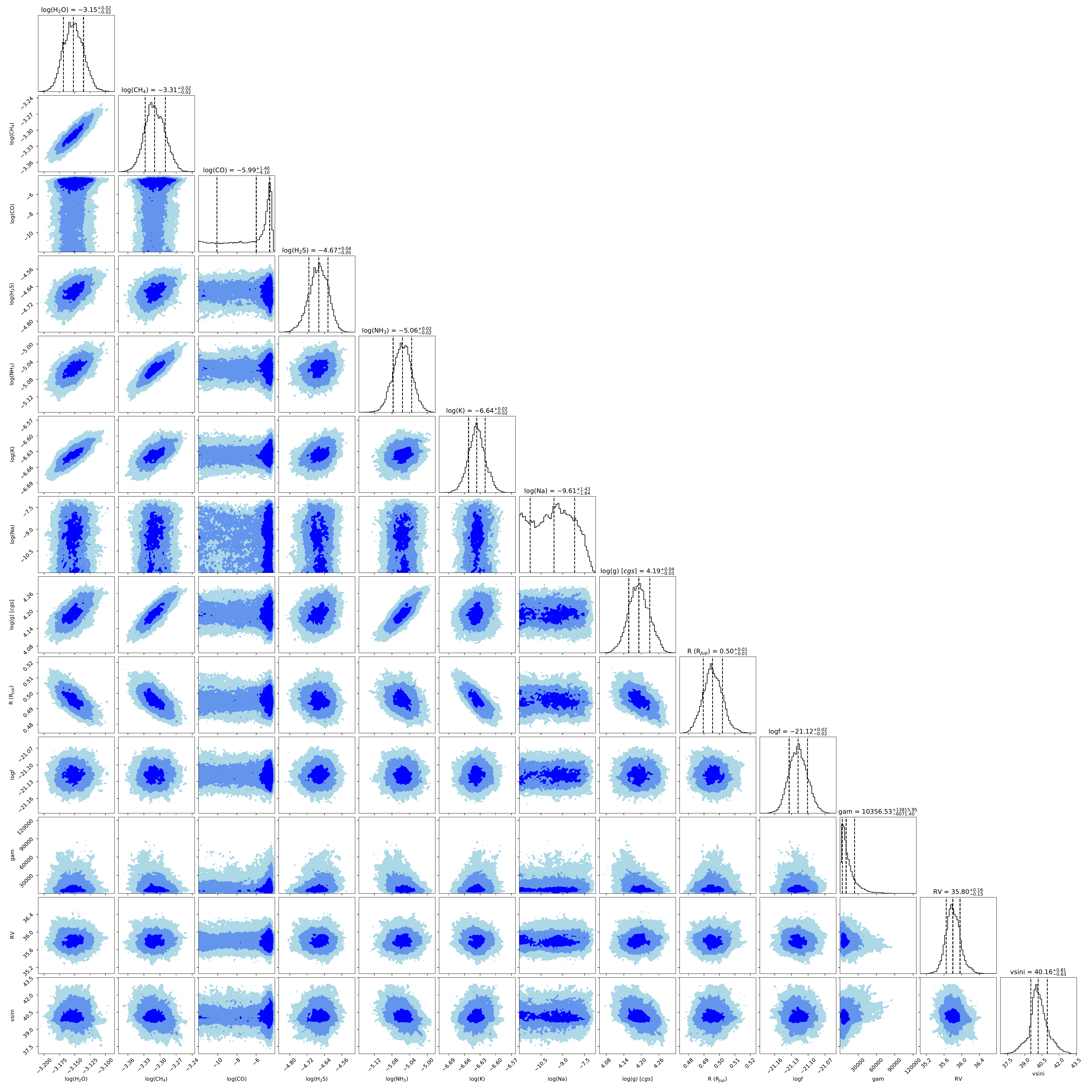}
  		\caption{Corner plot summarizing the posterior of our preferred FIRE retrieval for the listed parameters. Correlations between parameters are shown in the two-dimensional histograms. Marginalized posteriors for each parameter are shown along the diagonal, with the dashed lines indicating the 16th, 50th, and 84th percentiles. The median and $\pm 1\sigma$ values for each parameter are shown above the histograms. }
  		\label{fig:CornerPlot}
\end{figure*}

%

\vspace{5mm}



\bibliography{sample631}{}
\bibliographystyle{aasjournal}

\end{document}